\documentclass{aa}  
\usepackage{graphicx}
\usepackage{txfonts}
\usepackage[colorlinks=true, linkcolor=blue, citecolor=blue, urlcolor=gray]{hyperref}
\usepackage{xcolor}
\usepackage{listings}
\usepackage{braket}
\usepackage{physics}
\usepackage[hang,small,bf]{caption}
\usepackage[subrefformat=parens]{subcaption}
\newcommand{\Msun}{{\rm  M_{\odot}}}
\newcommand{\Zsun}{Z_{\odot}}

\begin{document} 
   \title{Evolution of galaxy attenuation curves driven by evolving dust mass and grain size distributions}
   \author{Kosei Matsumoto\inst{1}
         \fnmsep \thanks{\email{kosei.matsumoto@ugent.be}}
         \and
         Laura Sommovigo\inst{2}
         \and 
         Andrea Gebek \inst{1}
         \and
         Kentaro Nagamine\inst{4,5,6,7,8}
         \and
         Angelos Nersesian\inst{1,3}
         \and
         Maarten Baes\inst{1}
         \and
         Ilse De Looze\inst{1}
         \and
         Arjen van der Wel\inst{1}
         \and
         Rachel Somerville\inst{2}
         \and
         Leonard E. C. Romano\inst{9, 10,11}
         \and
         Rachel K. Cochrane\inst{12}
        }
   \authorrunning{Matsumoto et al.}
   \institute{
            Sterrenkundig Observatorium Department of Physics and Astronomy Universiteit Gent, Krijgslaan 281 S9, B-9000 Gent, Belgium
        \and
            Center for Computational Astrophysics, Flatiron Institute, 162 5th Avenue, New York, NY 10010, USA
        \and
            STAR Institute, Universit\'e de Li{\`e}ge, Quartier Agora, All\'ee du six Aout 19c, B-4000 Liege, Belgium
        \and
             Theoretical Astrophysics, Department of Earth and Space Science, Osaka University, 1-1 Machikaneyama, Toyonaka, Osaka 560-0043, Japan
        \and
            Theoretical Joint Research Project, Forefront Research Center, Graduate School of Science, Osaka University, 1-1 Machikaneyama, Toyonaka, Osaka 560-0043, Japan
        \and    
            Kavli IPMU (WPI), The University of Tokyo, 5-1-5 Kashiwanoha, Kashiwa, Chiba 277-8583, Japan
        \and
            Department of Physics and Astronomy, University of Nevada, Las Vegas, 4505 S. Maryland Pkwy, Las Vegas, NV 89154-4002, USA
        \and
            Nevada Center for Astrophysics, University of Nevada, Las Vegas, 4505 S. Maryland Pkwy, Las Vegas, NV 89154-4002, USA
        \and
            Universit\"{a}ts-Sternwarte, Fakult\"{a}t f\"{u}r Physik, Ludwig-Maximilians-Universit\"{a}t M\"{u}nchen, Scheinerstr. 1, D-81679 M\"{u}nchen, Germany
        \and
            Max-Planck-Institut f\"{u}r extraterrestrische Physik, Giessenbachstr. 1, D-85741 Garching, Germany
        \and
            Excellence Cluster ORIGINS, Boltzmannstr. 2, D-85748 Garching, Germany
        \and
            Jodrell Bank Centre for Astrophysics, University of Manchester, Oxford Road, Manchester M13 9PL, UK
             }

   \date{Received September 15, 1996; accepted March 16, 1997}


  \abstract
  {} 
  {We investigate the impacts of the evolution of dust mass and grain size distribution within a Milky Way-like (MW-like) galaxy simulation on the evolution of global attenuation curves, with a focus on the optical–ultraviolet (UV) slope and $2175$ \AA{} bump. Additionally, we discuss the contributions of star-dust geometry, scattering, and dust properties to the attenuation curves.} 
  {We performed post-processing dust radiative transfer using the SKIRT code based on a MW-like galaxy simulation. The hydrodynamic simulation was carried out with the GADGET4-OSAKA code, which models the evolution of grain size distributions.}
  {
    For lower inclination angles (i.e., closer to face-on), the attenuation curve flattens over time up to $t = 1$ Gyr, and then becomes progressively steeper. The steeper slope of the attenuation curve is caused by the interplay between scattering and the dust disk becoming more extended over time (i.e., changes in the star-dust geometry).
    At higher inclination angles, the effect of scattering is suppressed, and the attenuation curves slightly steepen over time due to the formation of small grains and the bias of observed UV emission toward old stars. 
    The $2175$ \AA{} bump becomes stronger on a timescale of $\sim250$ Myr due to the formation of small carbonaceous grains.  
    However, the bump strength is affected not only by the abundance of small grains but also by star-dust geometry. At higher $A_V$ or higher inclination angles, the bump strengths become weaker.
    These results may help interpret flatter attenuation curves and less prominent bumps in high-redshift galaxies.
    Furthermore, we find that variations in the star–dust geometry alter the amount of scattered photons escaping the galaxy, thereby driving the anti-correlation between the slope and $V$-band attenuation, $A_V$.
    The scatter in this relation arises from differences in dust optical depth along and perpendicular to the line of sight, reflecting differences in inclination and star-dust geometry. Additional contributions to the scatter come from variations in the grain size distribution and the fraction of obscured young stars.
    }
  {}




   \maketitle
%

\section{Introduction}
Dust in the interstellar medium (ISM) plays a crucial role in shaping the observable characteristics of galaxies. It absorbs and scatters stellar light across UV to optical wavelengths and re-emits energy in the mid- to far-infrared, significantly changing a galaxy's spectral energy distribution (SED). The galaxy attenuation curve, which quantifies the wavelength-dependent attenuation of stellar emission, is essential for deriving intrinsic galaxy spectra. As a result, accurate modeling of the attenuation curve is critical for reliably estimating key properties such as stellar mass, star formation rate, and age through SED fitting \citep[e.g., ][]{Walcher2011,Conroy2013}.

Although attenuation curves are often discussed alongside extinction curves \citep[see][for a review]{Salim_Narayanan_2020}, they are fundamentally different. Extinction curves are derived along the line of sight by comparing the spectrum of a reddened star to that of an unreddened star of the same type \citep{Cardelli1989, Fitzpatrick1999}. Extinction curves reflect only dust properties along the line of sight, such as grain size distribution and dust composition, and are useful for understanding local dust properties. However, extinction curves require resolving individual stars and have thus only been obtained for nearby galaxies like the MW, Small Magellanic Cloud (SMC), Large Magellanic Cloud (LMC), and M31 \citep[e.g., ][]{Koornneef&Code1981, Nandy1981, Prevot1984, Bouchet85, Cardelli1989, Mathis&Cardelli1992, Fitzpatrick1999, Valencic2004, Clayton2015, Gordon2023}.

In contrast, attenuation curves for galaxies depend not only on dust properties but also on the star-dust geometry and dust scattering, meaning they do not directly reflect the extinction curves of individual stars.
The attenuation curve can be estimated with several methods.
One method of measuring attenuation curves involves assessing the deviation of multiple Balmer emission line ratios from theoretically estimated intrinsic Balmer emission line ratios \citep{Calzetti1994, Calzetti1997b, Reddy2015, Battisti2016, Battisti2017a, Reddy2020,Sanders2024}. 
Using this method, \citet{Calzetti1997b} derived the Calzetti attenuation law ($A_\lambda \propto \lambda^{-0.75}$) based on the average attenuation of local starburst galaxies \citep{Calzetti1994, Calzetti2000}. 
In addition, numerous studies have fitted parameters of flexible attenuation laws via SED fitting \citep[$A_\lambda \propto \lambda^{-n}$,][]{Buat2012, Buat2018, Salim2018, Barisic2020, Nersesian2024, Nersesian2025, Boquien2022, Fisher2025, Battisti2022}. This approach was applied to high-redshift galaxies by \citet{Markov2023, Markov2024}, who adopted a more flexible dust attenuation law with additional free parameters.
\citet{Wild2011} employed the pair method, where they compared galaxies with the same type of intrinsic SED but different levels of dust attenuation to derive attenuation curves.
\citet{Salim2018} reported a wide diversity in both the slopes and the strengths of the $2175$ \AA{} bump among local galaxies. They also identified a correlation between the slope and the $V$-band attenuation ($A_V$), a trend that has been confirmed by several other observational studies \citep[e.g.,][]{Salmon2016, Leja2017, Decleir2019, Boquien2022}.
\citet{Battisti2017b} found that galaxies with lower inclination angles tend to exhibit steeper attenuation curves and weaker $2175$ \AA{} bumps \citep[see also][]{Wild2011, Barisic2020}.
While various studies have suggested that the attenuation curve depends on physical parameters such as metallicity, star formation rate, and stellar mass \citep[e.g.,][]{Shivaei2020, Battisti2020_2175Bump, Salim2018}, the fundamental drivers of these variations remain uncertain \citep{Salim_Narayanan_2020}.

The dust attenuation curves of galaxies have been extensively studied across a wide range of redshifts.
\citet{Battisti2022} reported that attenuation curves become steeper with increasing redshift up to $z < 3$, while the strength of the $2175$ \AA{} bump remains largely unchanged \citep[see also][]{Zeimann2015, Scoville2015}.
In contrast, recent James Webb Space Telescope (JWST) observations \citep{Markov2023, Markov2024, Markov2025} indicate that attenuation curves flatten and tend to lack prominent $2175$ \AA{} bumps at higher redshifts, suggesting a dust population dominated by larger grains in the early universe.
Similarly, \citet{Fisher2025} found flat attenuation curves in Lyman-break galaxies at $z = 6.5$–7.7.
On the other hand, \citet{Witstok2023} and \citet{Ormerod2025} identified strong $2175$~\AA{} bumps in galaxies at $z = 6$–7, indicating that small grains were already present in some early systems.
Collectively, these findings underscore the critical role of dust properties, particularly the grain size distribution, in shaping the attenuation curves of galaxies in the early universe.

To better interpret these observations, it is crucial to theoretically investigate physical mechanisms that shape attenuation curves.
Although radiative transfer calculations based on toy-model geometries have provided valuable insights into the underlying physics of attenuation curves \citep[e.g.,][]{Witt2000, Gordon2001, Chevallard2013, Seon2016, Nersesian2020, Lin2021},
those based on hydrodynamical simulations allow for the study of attenuation curves in the context of more realistic and self-consistent galaxy structures.
\citet{Trayford2020} found a strong relation between $A_V$ and the Near-UV slope based on attenuation curves in the EAGLE simulation \citep{Schaye2015}, which varies with galaxy inclination. \citet{Sommovigo2025} showed a correlation between $A_V$ and the star formation rate surface density ($\Sigma_\mathrm{SFR}$) in galaxies at $z=0.07$ from the TNG50 and TNG100 simulations \citep{Pillepich2018b,Pillepich2019}, suggesting that galaxies with higher star formation rate densities exhibit higher levels of dust attenuation and have flatter attenuation curves. \citet{Narayanan2018} demonstrated that star-dust geometry primarily drives the slopes of attenuation curves in galaxies at $z=0-6$ from the MUFASA simulations \citep{Dave2016}: galaxies with less obscured young stars show flatter attenuation curves. Both \citet{Narayanan2018} and \citet{Sommovigo2025} observed a decreasing trend in the strength of the $2175$ \AA{} bump with increasing $A_V$.
These studies highlight the importance of the star–dust geometry in shaping both the slope and the $2175$ \AA{} bump. However, they assume a fixed dust-to-metal mass ratio and a static dust model, where the grain size distribution and dust composition are not variable. This assumption may introduce uncertainties in the predicted dust distributions of galaxies.

Many studies have modeled physically motivated dust mass evolution alongside hydrodynamic simulations for galaxies \citep{Li2019,Graziani2020,Choban2022, Choban2024, Choban2025}, as well as the evolution of grain size distribution \citep{McKinnon2018, Aoyama2017, Aoyama2020, Hou2019, Romano2022Dust, Narayanan2023, Li2021, Dubois2024}. These models highlight the non-linear evolution of the dust-to-gas and dust-to-metal mass ratios, which vary according to the conditions of the ISM, where dust grains grow and are destroyed, as also inferred from several observational studies \citep[e.g.,][]{Remy-Ruyer2014A&A...563A..31R,DeVis2019,Galliano2021, Park2024, Algera2025}. These studies emphasize the importance of considering dust mass evolution.
\citet{Li2021} suggested that both grain size distribution and the graphite-to-silicate ratio influence the slope and $2175$ \AA{} bump in extinction curves. Similarly, \citet{Dubois2024} showed that SMC-like curves occur at low metallicity and MW-like curves occur at high metallicity.  However, the influence of evolving dust mass and grain size distribution, in conjunction with the evolving stellar populations and their geometry, on galaxy attenuation curves remains largely unexplored.

In this study, we explore the impact of dust mass and grain size evolution on attenuation curves using a MW-like galaxy simulation. We further examine the roles of scattering, star–dust geometry, and grain size distribution based on hydrodynamic simulations with dust evolution \citep{Matsumoto2024}.
This paper is structured as follows. Section~{\ref{method}} provides an overview of the physical models used in our hydrodynamic simulations with GADGET4-OSAKA and the methodology for post-processing radiative transfer with SKIRT. In Section~{\ref{sec: Evolution of the averaged extinction curve}}, we examine the time evolution of the average extinction curve of the MW-like galaxy simulation, focusing on how the evolution of the grain size distribution impacts the extinction curves. Section~{\ref{Sec: Evolution of the global attenuation curve}} explores the time evolution of the global attenuation curve, identifying the primary factors that influence the attenuation curve, including scattering, star-dust geometry, and grain size distribution. 
In Section~{\ref{Discussion}}, we compare our simulations with previous theoretical and observational studies, and present physical insights into dust attenuation curve variations.
Finally, in Section~{\ref{Conclusion}}, we present our conclusions.

%
%
\section{Method}\label{method}
\subsection{A MW-like galaxy simulation with GADGET4-OSAKA} 
\label{subsec: MW-like galaxy simulations}
We model an isolated MW-like galaxy using the methodology outlined in \citep{Matsumoto2024}. We summarize the methodology here, but refer the reader to \citet{Romano2022Dust, Matsumoto2024} for more details.
This simulation is performed using a smooth particle hydrodynamic (SPH) simulation code, GADGET4-OSAKA \citep{Romano2022Dust, Romano2022Mol,Oku2022,Oku2024}, a modified version of the GADGET4 code \citep{Springel2021}, incorporating the OSAKA feedback models \citep{Shimizu2019} and an evolutionary model for grain size distributions in the ISM \citep{Aoyama2020, Romano2022Dust}. 
The simulation follows the evolution of grain size distributions across $30$ bins ranging from $3.0 \times 10^{-4}$ to $10 \ \mathrm{\mu m}$ in a MW-like galaxy over $10$ Gyr\footnote{We used the snapshots at $t=0.05$, $0.1$, $0.25$, $0.5$, $1.0$, $2.0$, $3.0$, $4.0$, $5.0$, $6.0$, $7.0$, $8.0$, $9.0$, and $10.0$ Gyr in this paper.}, taking into account interstellar dust processes and dust production from asymptotic giant branch (AGB) stars and supernovae (SNe). 
The initial condition for the MW-like galaxy simulation is taken from the AGORA project\footnote{\url{https://sites.google.com/site/santacruzcomparisonproject/}} \citep{Kim2016}.
The components of the initial conditions of the MW-like galaxy simulation are summarized in Table~\ref{table:hydro parameters}.
The gas and stellar metallicity are initially set at $0.0001 \ \Zsun$, where $\Zsun = 0.013$ \citep{Asplund2009}, and increase to $0.7 \, \Zsun$ by $10$ Gyr. The gravitational softening length is set to 40 pc, and the minimum smoothing length is allowed to reach $0.1 \ \epsilon_\mathrm{grav}$. 

\begin{table}[h]
\caption{Components of the initial condition for the MW-like galaxy simulation. 
$M_\mathrm{total}$, $N_\mathrm{part}$, and $m_\mathrm{part}$ represent the total mass of each component, the number of particles for each component, and the mass of each particle, respectively. }  
\label{table:hydro parameters}      
\centering                          
\begin{tabular}{l | c c c }        
\hline\hline                 
Components & $M_\mathrm{total}$ ($\Msun$) & $N_\mathrm{part}$ & $m_\mathrm{part}$ ($\Msun$)\\    
\hline                        
   Gaseous disk  & $ 8.6\times10
   ^9$ & $1.0 \times 10^6$ & $ 8.6\times10^3$ \\
   Gaseous halo & $1.0 \times10^9$ & $4.0 \times 10^5$  & $ 4.0\times10^4$ \\
   Dark matter halo & $1.0\times10^{12} $  &  $1.0 \times 10^6$  & $ 1.0\times10^6$ \\
   Stellar disk & $ 3.4\times10^{10} $ & $1.0 \times 10^6$ & $ 3.4\times10^4$ \\
   Bulge stars & $4.3\times10^9 $ & $1.25 \times 10^5$ & $ 3.4\times10^4$ \\ 
\hline                                   
\end{tabular}
\end{table}

Star formation occurs stochastically, governed by a star formation efficiency of $\epsilon_* = 0.05$, as per the Kennicutt-Schmidt law \citep{Kennicutt1998}. Star formation occurs when the density of gas particles exceeds $n_\mathrm{H} > 20 \ \mathrm{cm^{-3}}$ and the temperature of the gas is below $T_\mathrm{gas} < 10^4$ K.
In the OSAKA feedback model, the feedback processes of type Ia SNe, type II SNe, and AGB stars affect the physical conditions of the ISM within the galaxy \citep{Shimizu2019} via thermal and kinetic feedback processes. Simultaneously, SNe and AGB stars eject metal elements and dust. The ejected metal mass is calculated using the CELib library \citep{Saitoh2017}, and 10\% of the ejected metals are assumed to condense into dust grains \citep{Asano2013}.
We employed the GRACKLE-3 radiative cooling library \citep{Smith2017}\footnote{\url{https://grackle.readthedocs.org/}}, which solves the non-equilibrium primordial chemical network, taking into account metal cooling as well as photo-heating and photo-ionization due to the UV background \citep{Haardt&Madau2012}.

Our simulations lack the necessary spatial resolution to accurately resolve cold and dense gas ($T_\mathrm{gas} < 50$ K and $n_\mathrm{H} > 10^3$ cm$^{-3}$). Therefore, we implemented a two-phase ISM subgrid model for gas particles \citep{Romano2022Dust}, assuming that each gas particle contains both dense and diffuse gas phases. The mass fraction of the dense gas phase is given by
\begin{equation}
    f_\mathrm{dense} = \mathrm{min}\, \Bigg( \alpha \, \frac{n_\mathrm{H}}{1.0 \ \mathrm{cm}^{-3}}, 1.0\Bigg),\label{eq: dense gas fraction}
\end{equation}
for which we adopted $\alpha =0.12$.
This equation is based on the observation that regions with higher gas concentrations in the ISM typically host a larger fraction of dense molecular gas clouds \citep{Gnedin2011, Gnedin2014}.
\citet{Romano2022Dust} calibrated the parameter $\alpha$ to reproduce the observed mean molecular gas fraction for the MW-mass galaxies with stellar masses around $10^{10}$ $\Msun$ \citep{Catinella2018}. While the temperature and density of the dense gas phase were assumed to be constant ($T_\mathrm{dense} = 50$~K and $n_\mathrm{H, \ dense} = 10^3$~cm$^{-3}$), the temperature and density of the diffuse gas phase in each gas particle were determined by considering the conservation of internal energy. This subgrid model enabled us to separately treat dust processes in the diffuse and dense ISM, accounting for their distinct properties within each gas particle.

We take into account various interstellar dust processes in both the diffuse and dense gas phases. In the ISM, shattering breaks down larger grains into smaller ones, while coagulation leads to the growth of small grains into larger ones. Shattering and coagulation are assumed to occur through grain–grain collisions in the dense and diffuse gas phases, respectively. Both processes are influenced by the grain size and velocities, with the grain velocity determined by its coupling to turbulent motion, which follows a Kolmogorov spectrum \citep{Kolmogorov1941} and reflects the local physical conditions.
Additionally, metal accretion onto dust grains predominantly takes place in the dense gas phase, causing an increase in the size and mass of the grains. The accretion rate is determined by the collision frequency between gas-phase metals and dust, and it is parameterized by the dust-to-metal ratio, grain size, temperature, and metallicity. On the other hand, sputtering, which occurs in the diffuse gas phase, reduces the size and mass of each grain. The sputtering rate in hot gas is based on \citet{Tsai1995}, and it is also influenced by the sweeping rate of supernova shocks that contribute to the destruction of dust.
The chemical composition of grains is not explicitly included in the dust evolution model of the GADGET4-OSAKA, and thus, we model the dust composition of each dust grain in post-processing (see Section \ref{subsec:SKIRT}).
We refer to \citet{Aoyama2020} and \citet{Romano2022Dust} for more details on the treatment of dust processes in the simulation.

\subsection{Post-processing radiative transfer with SKIRT}
\label{subsec:SKIRT}
We performed radiative transfer in post-processing based on the snapshots of the MW-like galaxy simulation with the radiative transfer code, SKIRT\footnote{\url{www.skirt.ugent.be.}}
\citep[version 9.0;][]{Baes2011,Camps&Baes2015,Camps2020SKIRT9}, to generate attenuation curves. SKIRT has been widely used for dust radiative transfer calculations \citep{Camps2018, Liang2018, Ma2019, Parsotan2021, Cochrane2019,Cochrane2022,Cochrane2023a,Cochrane2023b,Cochrane2024,Camps2022, Kapoor2021,Kapoor2023,Kapoor2024, Trcka2022,Baes2024a, Baes2024b, Gebek2024, Gebek2025, Sommovigo2025}, although it is a generic radiative transfer code with capabilities to handle many radiative processes relevant to dust and gas \citep[e.g.,][]{Camps2021, Gebek2023, Matsumoto2023, Bert2023, Bert2024a, Bert2024b}. 

First, we used the \citet{Bruzual&Charlot2003} templates with the \citet{Chabrier2003} initial mass function for single stellar populations. The intrinsic SED of each star particle is determined based on its stellar age, metallicity, and initial stellar mass. 
Under the initial conditions of the hydrodynamic simulations, the disk and bulge star particles are pre-assigned to make the galaxy dynamically stable. This means that some stars already exist at the simulation time of $t=0$ Gyr. Since the simulations run until $t=10$ Gyr, and we assume the final galaxy age to be comparable to the cosmic age ($13.8$ Gyr), the disk and bulge stars are assumed to have formed between $t=-3.8$ and $t=0$ Gyr.
Therefore, in post-processing, we randomly assigned their ages to obtain a constant star formation history within the $3.8$ Gyr interval prior to $t=0$ Gyr. We note that the pre-assignment of stellar age did not significantly impact the evolution of the galaxy attenuation curves on Gyr timescales, as the stellar emission is primarily dominated by the stars formed during the simulation. Here, we assume that star particles are point sources with a fixed size of $0.1$ pc. This assumption is reasonable because star particles are treated as collisionless particles in the GADGET4-Osaka simulation, and their assumed size does not significantly affect the galaxy SED.

Second, for modeling dust composition, we decomposed the dust grains in each gas particle into silicate, carbonaceous dust, and PAHs in post-processing. 
Here, we follow the method of decomposition of dust grains into the various chemical compositions described in \citet{Matsumoto2024}.
First, we assigned larger dust grains (with sizes $a > 0.0013$ $\mathrm{\mu}$m) as silicate and carbonaceous dust grains according to the abundances of Si and C in each gas particle \citep{Hirashita&Murga2020}.
We distinguish between small and large grains using a grain size boundary of $a = 0.0013$ $\mathrm{\mu}$m, which is derived from the definition of the maximum size of PAHs as described in \citet{Draine2001DustsizeDefinition} and \citet{Draine2007PAHDefinition}.
Small grains (with sizes $a \leq 0.0013$ $\mathrm{\mu}$m) were designated as PAHs and non-aromatic carbonaceous grains. The mass fraction of non-aromatic carbonaceous grains is determined by the dense gas fraction in each gas particle ($f_\mathrm{aroma} = 1 - f_\mathrm{dense}$; \citealt{Hirashita&Murga2020}). Based on this assumption, the majority of small grains with $a \leq 0.0013$ $\mathrm{\mu}$m were assigned to PAHs, as these small grains are primarily produced in the diffuse ISM.
Finally, to decompose PAHs into neutral (PAH$^0$) and ionized (PAH$^+$) forms, we used the ionization fraction derived from the 'standard' ionization model in \citet{Draine2021} and \citet{Hensley2023}. As a result, PAH$^+$ are primarily assigned to the smallest grains ($a \sim 0.0003$ $\mathrm{\mu}$m).

\begin{figure}[]
    \includegraphics[width=0.5\textwidth]{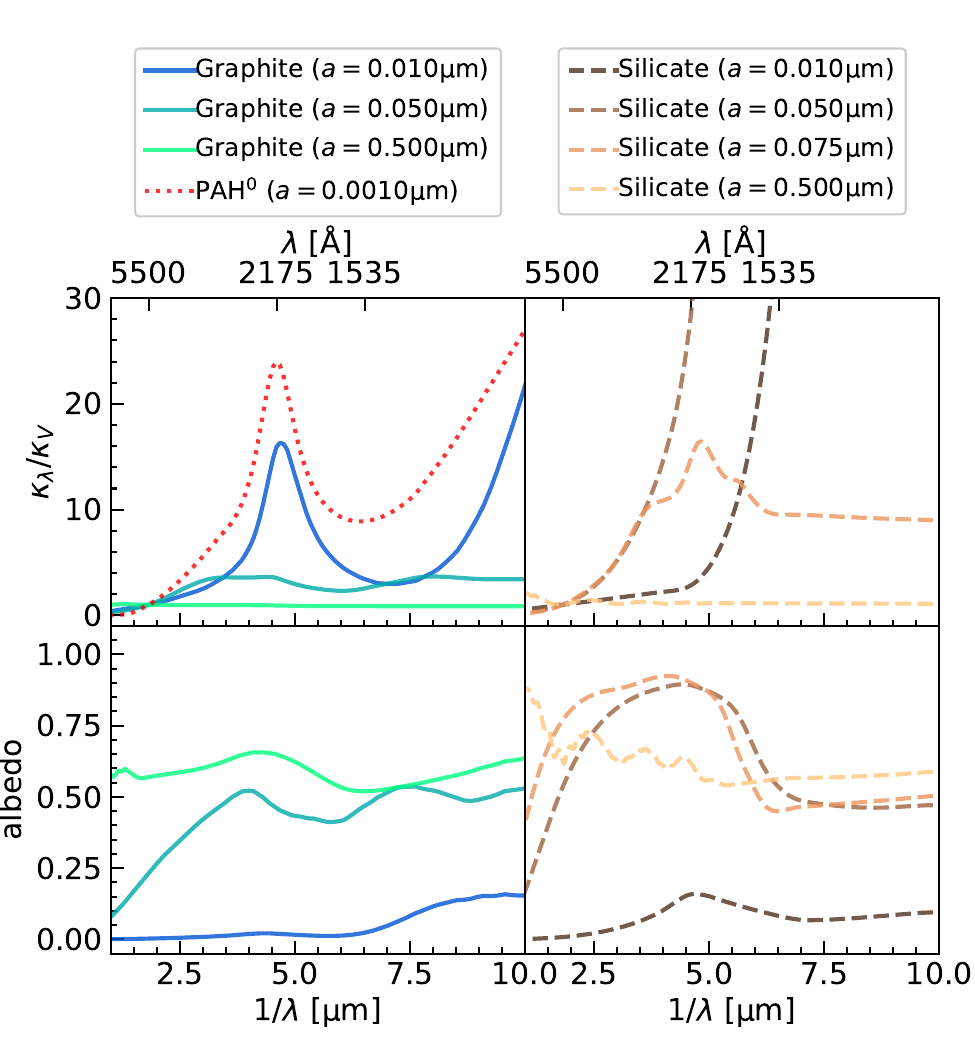}
    \caption{Extinction curves for carbonaceous and silicate grains with a given dust grain size (upper left and right, respectively). The extinction curve is normalized by the opacity at the $V$-band wavelength ($\lambda=5500$ \AA{}). The lower panels show the wavelength dependence of albedo, which is the fraction of scattering against extinction, of carbonaceous and silicate grains (lower left and right, respectively).}
    \label{fig: Extinction curve and albedo}
\end{figure}

The opacity and calorimetric properties of PAHs, silicate, and non-aromatic carbonaceous grains (assumed to be graphite) follow the prescription of \citet{Draine&Li2007}. 
The upper panels of Fig.~\ref{fig: Extinction curve and albedo} show the extinction curve for carbonaceous and silicate grains with a given grain size (upper left and right, respectively) taken from \citet{Draine&Li2007}. The extinction curve is derived from the wavelength dependence of the dust grain opacity, $\kappa_\lambda$, and is normalized at the $V$-band wavelength ($\lambda = 5500$ \AA{}).
As the grain size increases, the extinction curves for silicate and graphite grains become flatter. The $2175$ \AA{} bump strength of graphite grains develops when the grain size is lower than $0.05$ $\mathrm{\mu m}$. PAH shows a steep extinction curve with strong $2175$ \AA{} bumps compared to the extinction curve for graphite grains.
We note that the extinction curves for graphite grains with a size of $10^{-4}$ $\mathrm{\mu m}$ are nearly identical to those for grains of size $10^{-3}$ $\mathrm{\mu m}$.

The lower panels of Fig.~\ref{fig: Extinction curve and albedo} show the wavelength dependence of the albedo, which is the fraction of scattering against total extinction, for carbonaceous and silicate grains with a given dust grain size (lower left and right, respectively). The albedo of graphite increases with grain size in the optical-to-UV wavelength range, while smaller grains exhibit lower albedo at longer wavelengths. Smaller silicate grains show higher albedo around $2000$ \AA{}, whereas larger silicate grains, up to a size of 0.2 $\mathrm{\mu m}$, have higher albedo in the optical wavelengths. For even larger silicate grains, the albedo tends to decrease at optical wavelengths. 
Thus, larger grains are responsible for scattering at optical wavelengths. Additionally, silicate grains tend to cause more scattering compared to graphite.
We note that graphite and silicate grains with sizes smaller than $10^{-4}$ $\mathrm{\mu m}$, as well as PAHs, do not contribute to scattering, resulting in an albedo of 0.0 across all wavelengths.

Given the dust and stellar distributions from the GADGET4-OSAKA simulation and dust properties from \citet{Draine&Li2007}, we performed post-processing radiative transfer calculations with SKIRT, incorporating spatially varying grain size distributions and chemical composition of the grains.
For the SKIRT simulations, the dust distribution of SPH particles, smoothed with the cubic kernel function, is resampled onto an octree grid with refinement based on a dust mass fraction threshold of $10^{-6}$ relative to the total dust mass.
We show the time evolution of dust mass and star formation history of the MW-like galaxy simulation in Appendix~\ref{app: Global properties of the MW galaxy simulation}.
For generating synthetic images with SKIRT, the field of view with ($40$ kpc)$^2$ is composed of spaxels with a pixel size of 50 pc and a spectral resolution of $R=\frac{ \lambda }{\Delta \lambda}=500$ in the wavelength ranges from $1000$ to $10000$ \AA{}. We set the inclination angles from $0^\circ$ to $90^\circ$ in increments of $10^\circ$ (from face-on to edge-on views).

\subsection{Modeling global extinction and attenuation curves} \label{subsec:Parametrization of extinction and attenuation curves}
We assume spherical grains with material density, $\rho$, such that the grain mass, $m(a)$, and radius, $a$, are related as $m(a)=\frac{4}{3}\pi a^3 \rho$. The grain size distribution, denoted as $\frac{dn(a)}{da}$, is defined so that $n(a)$ is the number distribution of grains as a function of grain radius $a$.
Using the global grain size distribution across the entire galaxy, we calculated the average extinction curve for each dust component, normalized by the total extinction at the $V$-band ($\lambda = 5500$~\AA{}), as follows:
\begin{equation}
\frac{\tau_\lambda^{\mathrm{comp}}}{\tau_V^{\mathrm{total}}}
= \frac{
\int \kappa_\lambda^{\mathrm{comp}}(a) \, m^{\mathrm{comp}}(a)\, \left( \frac{dn(a)}{da} \right)_{\mathrm{comp}} da
}{
\sum\limits_{\mathrm{comp}} \int \kappa_V^{\mathrm{comp}}(a) \, m^{\mathrm{comp}}(a) \, \left( \frac{dn(a)}{da} \right)_{\mathrm{comp}} da
},
\label{eq:averaged_extinction_curve}
\end{equation}
where $\kappa_\lambda^\mathrm{comp}(a)$ represents the extinction opacity for each dust component as a function of grain size at a given wavelength. 
The total average extinction curve is given by $\tau_\lambda^\mathrm{total}/\tau_V^\mathrm{total} = 
\sum\limits_{\mathrm{comp}}\tau_\lambda^\mathrm{comp}/\tau_V^\mathrm{total}$ and influenced solely by the grain size distribution and the chemical composition of the dust.

On the other hand, the attenuation curve is estimated based on the observational properties given by
\begin{equation}
    A_\lambda = -2.5 \log \Bigg{(}\frac{F_\lambda^\mathrm{observed}}{F_\lambda^\mathrm{intrinsic}}\Bigg{)},
    \label{eq: attenuation}
\end{equation}
where $F_\lambda^\mathrm{observed}$ and $F_\lambda^\mathrm{intrinsic}$ are the observed and intrinsic flux densities of the galaxy, respectively. 
We note that the attenuation is not an additive property, as it is influenced not only by the dust properties but also by scattering and star-dust geometry.

We parameterized the extinction and attenuation curves using the UV-to-optical slope and $2175$ \AA{} bump strength (hereafter referred to as the slope and bump, respectively).
The slope is defined as 
\begin{equation} \label{eq: definition of slope}
S=\frac{A_{FUV}}{A_V},
\end{equation}
where $A_{FUV}$ and $A_V$ represent the attenuation at $FUV$- and $V$-bands ($\lambda=1535$ and $5500$ \AA{}), respectively.
To derive the bump strength, we fitted the extinction and attenuation curves by the following function from \citet{Li2008}, 
\begin{equation}\label{eq: Li08}
\begin{split}   
\frac{A^\mathrm{fit}_{\lambda}}{A_{V}}
& = \frac{c_1}{(\lambda/0.08)^{c_2}+(0.08/\lambda)^{c_2} +c_3} \\ 
& +  \frac{233[1-c_1/(0.145^{-c_2}+0.145^{c_2}+c_3)-c_4/4.60]}{(\lambda/0.046)^2+(0.046/\lambda)^2+90} \\ 
& + \frac{c_4}{(\lambda/0.2175)^2+(0.2175/\lambda)^2-1.95},
\end{split}
\end{equation}
where $c_1$, $c_2$, $c_3$ and $c_4$ are dimensionless parameters and $\lambda$ is the wavelength in $\mu \rm{m}$ units \citep[][]{Markov2023, Markov2024, Sommovigo2025}. The way these parameters are reflected in the attenuation curves is discussed in \citet{Sommovigo2025}. Consequently, the bump strength is defined as
\begin{equation} \label{eq: definition of bump}
    B = \frac{A_{2175\AA}-A_{2175\AA, \ 0}}{A_{2175\AA}},
\end{equation}
where $A_{2175\AA, \ 0}$ is the baseline of the attenuation at 2175 \AA{}, calculated by $A^\mathrm{fit}_{2175\AA} (c_1,c_2,c_3,c_4=0)$. For the extinction curves, $A_\lambda$ in the equation~(\ref{eq: definition of slope}), (\ref{eq: Li08}), and (\ref{eq: definition of bump}) is replaced by $\tau_\lambda$.


%
%
\section{Time-Evolution of the average extinction curve}\label{sec: Evolution of the averaged extinction curve}

\begin{figure}[htbp]     \includegraphics[width=0.5\textwidth]{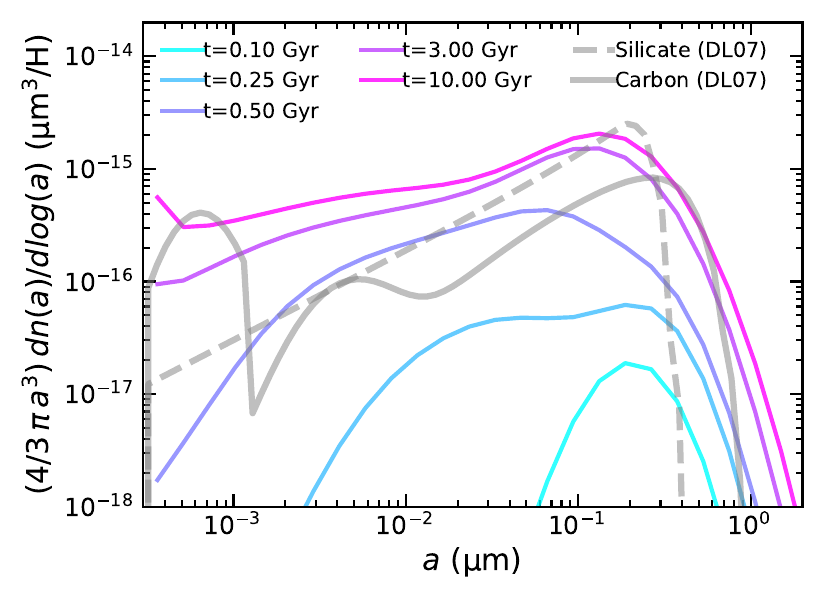}
    \caption{Time evolution of the global grain size distribution in our MW-like galaxy simulation. The solid and dashed gray lines correspond to the grain size distributions of carbonaceous and silicate grains in the static dust model from \citet{Draine&Li2007}. }%
    \label{fig: the evolution of the grain size distribution}
\end{figure}

First, we briefly introduce the global evolution of the grain size distribution in the MW-like galaxy simulation shown in Fig.~\ref{fig: the evolution of the grain size distribution}. Full details of the dust evolution scenario are described in \citet{Romano2022Dust} and \citet{Matsumoto2024}.
The solid lines in Fig.~\ref{fig: the evolution of the grain size distribution} represent the grain size distribution of the MW-like galaxy at various simulation times ($t=0.1$, $0.25$, $0.5$, $3.0$, and $10.0$ Gyr). 
We note that, in our model, the evolution of the grain size distributions for silicate and carbonaceous grains does not differ significantly.
At $t=0.1$ Gyr, only large dust grains ($a\sim0.2$ $\mathrm{\mu m}$) are produced by type II SNe in the galaxy. Then, shattering converts the large grains to small ones in the diffuse ISM, and accretion helps dust grains to increase their sizes. Thus, small grains ($a<0.1$ $\mathrm{\mu m}$) gradually form in the galaxy up to $t=0.5$ Gyr. Later, small grains start being coagulated to large grains ($a>0.1$ $\mathrm{\mu m}$) in the dense ISM, while shattering keeps producing small grains in the diffuse ISM.

\begin{figure}[htbp]
    \centering
    \includegraphics[width=0.5\textwidth]{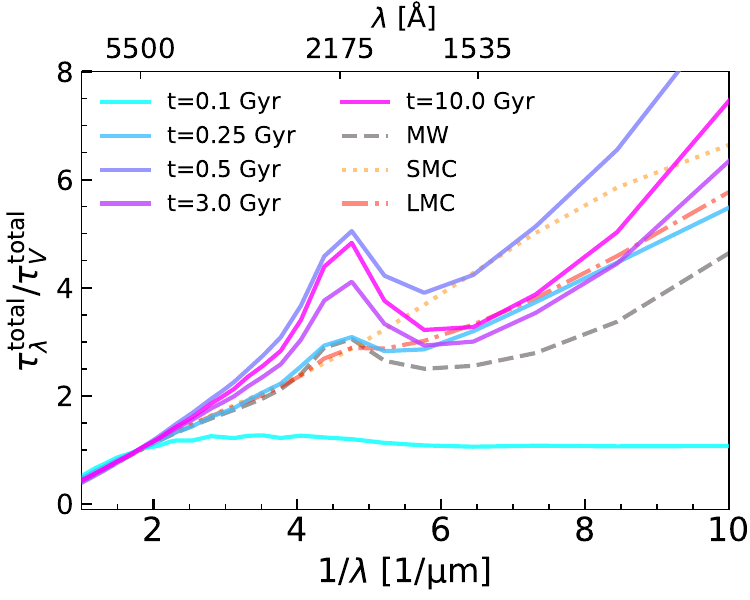}
    \caption{The time evolution of the average extinction curve of the MW-like galaxy simulation indicated by colored solid lines. The extinction curve is estimated by Eq.~\ref{eq:averaged_extinction_curve} and normalized by the opacity at the $V$-band wavelength ($\lambda=5500$ \AA{}).
    The dashed gray, dotted orange, and dash-dotted red lines represent the extinction curves of the MW, SMC, and LMC, respectively \citep{Pei1992, Li2008}.}%
    \label{fig: the evolution of the extinction curve}
\end{figure}

Figure~\ref{fig: the evolution of the extinction curve} shows the evolution of the average extinction curve of the MW-like galaxy simulation, derived using Eq.~\ref{eq:averaged_extinction_curve}. The extinction curve purely reflects the global grain size distribution and chemical composition of dust grains within the MW-like galaxy simulation. 
Although it is impossible to observationally obtain the average extinction curve, it is essential to theoretically explore how the evolution of grain size distributions is reflected in the average extinction curve.
The average extinction curve becomes progressively steeper up to $t=0.5$ Gyr along with a development of the $2175 \ \AA$ bump, driven by the production of smaller grains through shattering and accretion. After that, the extinction curve becomes gradually flatter up to $t = 3$ Gyr because small grains are coagulated. Later, the extinction slightly steepens until $t = 10$ Gyr because of the formation of very small grains including PAH through shattering. 
At all times, the evolution of the $2175 \ \AA$ bump strength is coherent with that of the slope. The detailed evolution of the slope and the $2175$~\AA{} bump strength is presented in Sections~\ref{sec: Time evolution of the slope at various inclination angles} and \ref{sec: Time evolution of 2175 bump at various inclination angles}.

\begin{figure}[]
    \centering
    \includegraphics[width=0.5\textwidth]{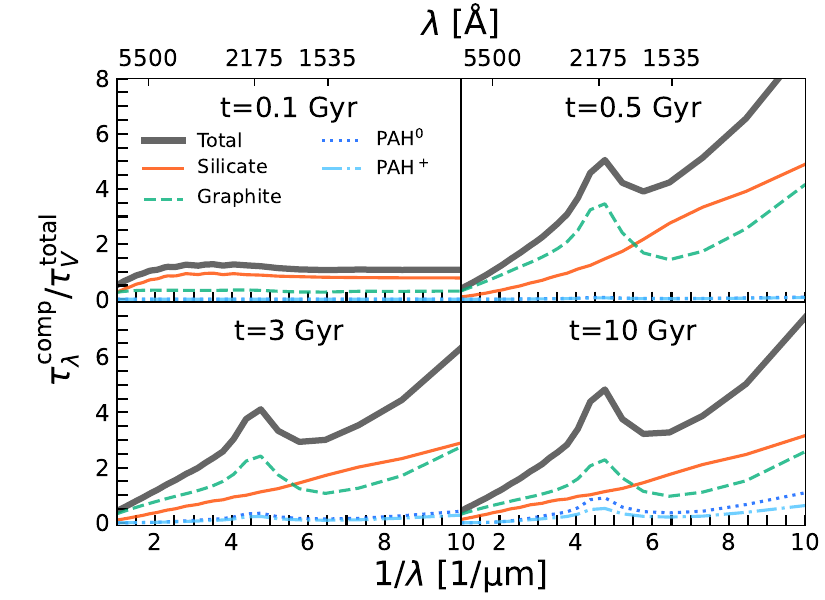}
    \caption{The contribution of each dust component to the average extinction curve of the MW-like galaxy simulation at various times ($t=0.1$, $0.5$, $3.0$, and $10.0$ Gyr). The black thick line is the total average extinction curve, while the orange solid, green dashed, blue dotted, and light blue dash-dotted thin lines represent the contributions of silicate, graphite, PAH$^0$, and PAH$^+$, respectively. At $t = 0.1$ Gyr, the extinction curve is dominated by silicate grains across all wavelengths. However, at later times, carbonaceous grains become the dominant contributors to extinction at wavelengths longer than $2000$~\AA{}.}%
    \label{fig: the evolution of the extinction curve of each component}
\end{figure}

Figure~\ref{fig: the evolution of the extinction curve of each component} shows the contribution of each dust component to the average extinction curve of the MW-like galaxy simulation at various times. 
At $t=0.1$ Gyr, the extinction curve is mainly driven by silicate at all wavelengths. At this time, gas-phase silicon and carbon are produced by SNe II. After $t=0.5$ Gyr, the extinction curve at shorter wavelengths is determined by graphite, since gas-phase carbon is produced by SNe Ia \citep{Saitoh2017}.  
Also, the $2175$ \AA{} bump becomes prominent due to the formation of small graphite grains ($a<0.05$ $\mathrm{\mu m}$).  
At $t=3$ Gyr, the extinction curve flattens because of the coagulation of small grains.
Later, the bump becomes stronger, and the slope steepens thanks to the production of very small grains, especially PAHs. 
Overall, silicate grains are responsible for the extinction curve at FUV wavelengths at all times, while carbonaceous grains become the dominant contributors to the extinction curve at wavelengths longer than $2000$~\AA{}, similar to previous studies \citep{Weingartner2001, Jones2013, Dubois2024}.

\section{Time evolution of the global attenuation curve}
\label{Sec: Evolution of the global attenuation curve}

\begin{figure*}[]
    \centering
    \includegraphics[width=0.8\textwidth]{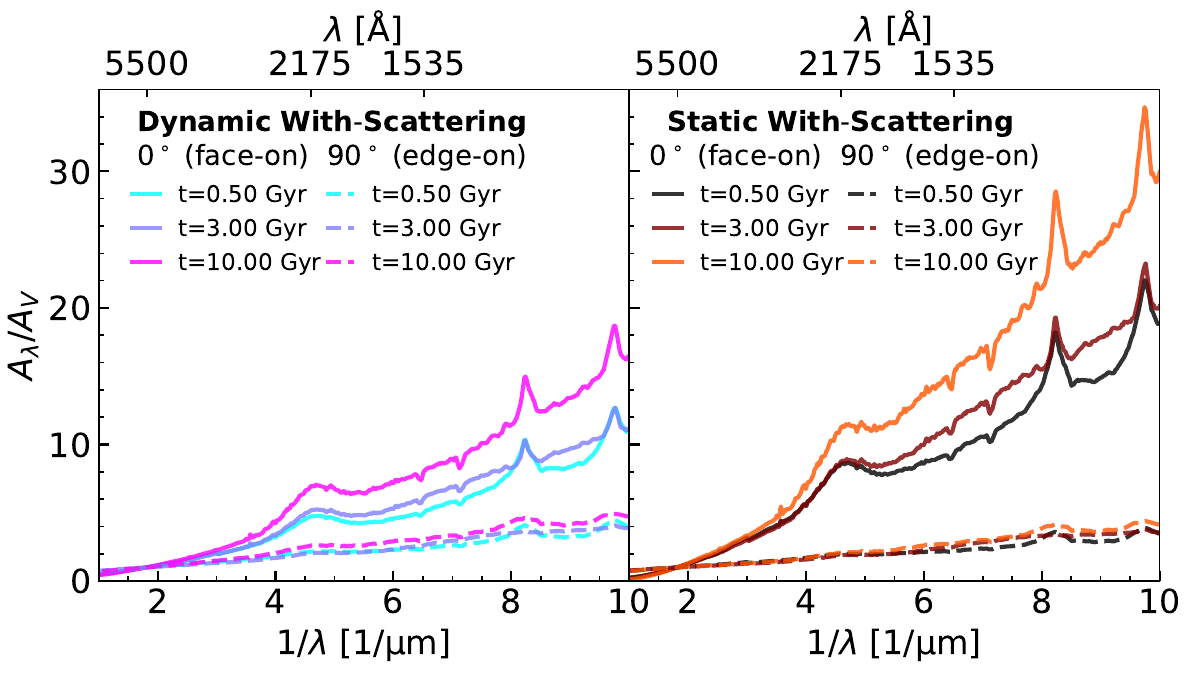}
    \caption{Time evolution of the global attenuation curve for the MW-like galaxy simulation using the \textit{Static With-Scattering} and \textit{Dynamic With-Scattering} models shown in left and right panels, respectively. The solid and dashed lines represent inclination angles of $0^\circ$ and $90^\circ$ (i.e., face-on and edge-on view, respectively).
    The attenuation curve is estimated using Eq.~\ref{eq: attenuation} and normalized by the attenuation at the $V$-band wavelength ($\lambda=5500$ \AA{}).
    }%
    \label{fig: the evolution of the attenuation curves}
\end{figure*}

The attenuation curve is influenced by various factors such as (i) scattering \citep{Chevallard2013}, (ii) star-dust geometries \citep{Seon2016,Narayanan2018}, and (iii) dust properties including grain size distribution and chemical composition.
Ultimately, these factors can be reflected in the relation between the attenuation curve and global galaxy properties and inclination effects \citep{Trayford2020,Sommovigo2025}.
However, which factor shapes the attenuation curve is not understood well. In order to examine and isolate these competing factors, we generated attenuation curves using two dust models: the first is a dynamic dust model utilizing dust grain size distribution and chemical composition from our GADGET4-OSAKA simulation (the details of the decomposition are described in Section~\ref{subsec:SKIRT}); the second is a static dust model with a fixed grain size distribution and composition based on the MW model from \citet{Draine&Li2007}. Although the second model accounts for dust mass evolution, it does not use the grain size distribution from the GADGET4-OSAKA simulation. A comparison of the grain size distributions of the two dust models is shown in Fig.~\ref{fig: the evolution of the grain size distribution}. 
Moreover, we generated attenuation curves considering scattering explicitly and handling scattering as absorption (i.e., without scattering) to probe the impact of scattering on attenuation curves.
We note that, even when distinguishing between these models, the scattering processes still depend on the star–dust geometry and dust properties.
Consequently, four types of attenuation curves are created and summarized below:
\begin{enumerate}
  \item \textbf{\textit{Static No-Scattering}:}
The attenuation curves are derived from the static dust model, excluding scattering effects. The attenuation curves are affected only by the star-dust geometry.

  \item \textbf{\textit{Static With-Scattering}:}
  The attenuation curves are derived using the static dust model, and scattering is explicitly taken into account.
  The attenuation curves are affected by the star-dust geometry and scattering. 
  
  \item \textbf{\textit{Dynamic No-Scattering}:}
    The attenuation curves are derived using the dynamic dust model, excluding scattering effects. The attenuation curves are driven by the star-dust geometry and dust properties including grain size distribution and dust composition.

  \item \textbf{\textit{Dynamic With-Scattering}:}
    The attenuation curves are derived using the dynamic dust model, and scattering is explicitly taken into account. The attenuation curves are determined by a combination of star-dust geometry, scattering, and evolving dust grain properties. This is the most physically comprehensive model of the four considered here.
\end{enumerate}


First, we present an overview of the time evolution of the attenuation curve and its dependence on different inclination angles. 
Figure~\ref{fig: the evolution of the attenuation curves} shows the evolution of the attenuation curves of our MW-like galaxy simulation with the \textit{Dynamic With-Scattering} and \textit{Static With-Scattering} models.
In both models, the attenuation curves for an inclination angle of $90\,^\circ$ (i.e., edge-on view) are flatter than those at $0\,^\circ$ (i.e., face-on view), and the $2175$~\AA{} bumps become less pronounced.
The attenuation curves for the \textit{Static With-Scattering} model become progressively steeper over time in the face-on view, despite assuming a constant grain size distribution and dust composition, while the slope remains nearly constant in the edge-on view. The \textit{Dynamic With-Scattering} model shows a similar evolutionary trend.
We caution that spikes  in the attenuation curves are caused by star–dust geometry effects. Ly$\alpha$ and Ly$\beta$ absorption lines from older stellar populations, as modeled in the \citet{Bruzual&Charlot2003} SSPs, appear as convex-shaped features in the attenuation curves. In contrast, absorption lines originating exclusively from younger stars present as concave-shaped features.
We quantify the time evolution of the slope and bump strength more accurately in the following sections where we discuss the following factors: (i) scattering, (ii) star-dust geometry, and (iii) dust properties by examining the difference in the slope and bump strength of the attenuation curves from the four different models (\textit{Dynamic With-Scattering, Dynamic No-Scattering, Static With-Scattering}, and \textit{Static No-Scattering}).

\subsection{Time evolution of the slope at various inclination angles}
\label{sec: Time evolution of the slope at various inclination angles}

\begin{figure*}[]
    \centering
    \includegraphics[width=0.95\textwidth]{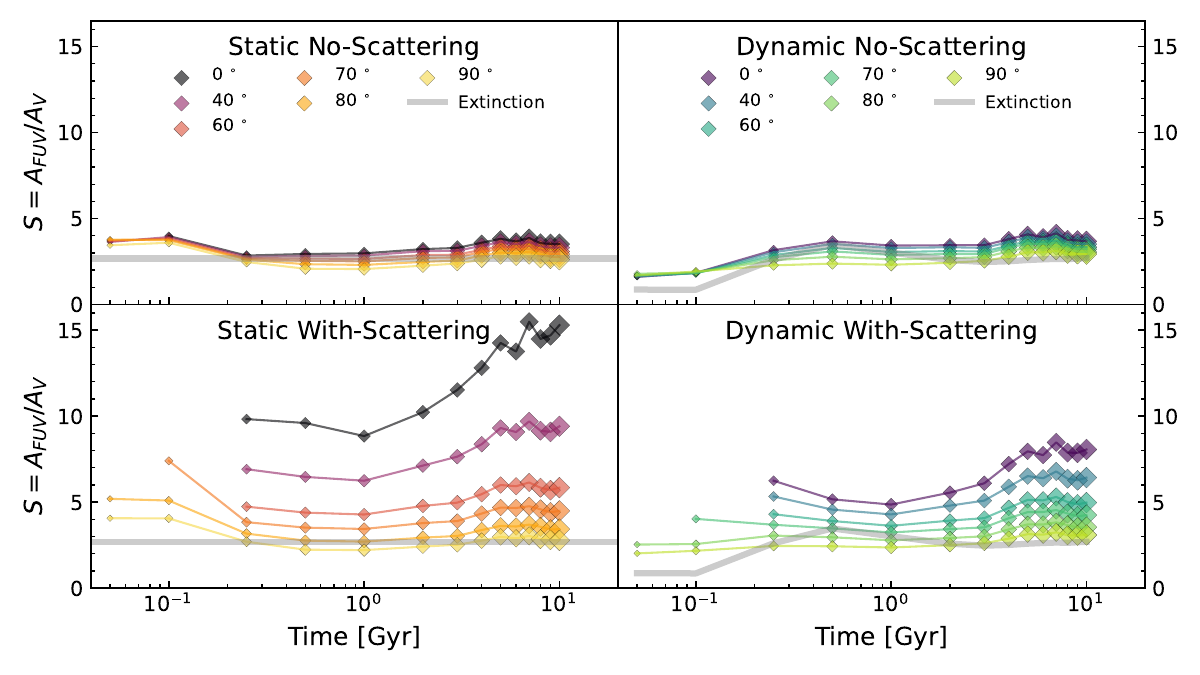}
    \caption{Time evolution of the slope for the global attenuation curves from the four different models (\textit{Dynamic, Dynamic No-Scattering, Static}, and \textit{Static No-Scattering}) of the MW-like galaxy simulation. The inclination angle is color-coded in each panel. Here, all models with $A_\mathrm{v}<0.01$ are excluded.
    The gray thick lines represent the time evolution of the slope of the average extinction curves for the static and dynamic dust models in the left and right panels, respectively. The value of slope for the static dust model corresponds that of the MW \citep{Pei1992, Li2008}.}%
    \label{fig: the time evolution of slopes}
\end{figure*}
\begin{figure*}[]
    \centering
    \includegraphics[width=0.95\textwidth]{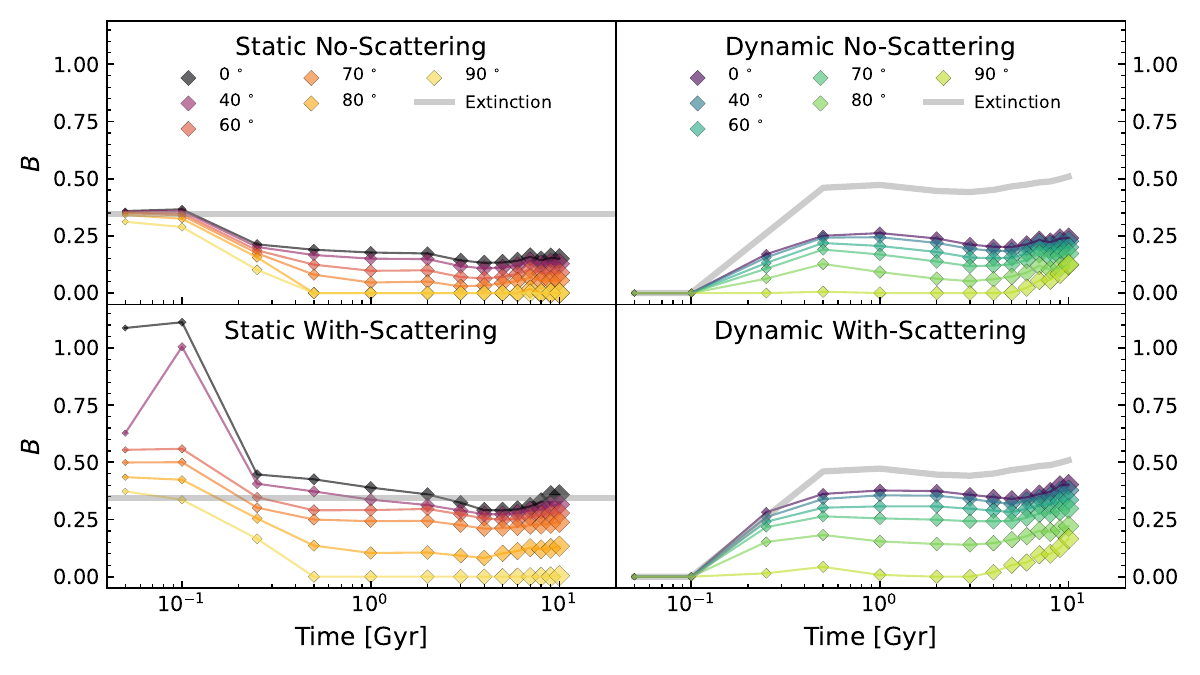}
    \caption{Same as Fig.~\ref{fig: the time evolution of slopes}, but time evolution of the $2175$ \AA{} bump strength.}%
    \label{fig: the time evolution of bump}
\end{figure*}

We examine the time evolution of the slope of the global attenuation curve of the MW-like galaxy simulation by comparing that of the extinction curve. 
Figure~\ref{fig: the time evolution of slopes} presents the time evolution of the slope of the global attenuation curve for the four different models of the MW-like galaxy simulation at various inclination angles. 
Firstly, in the \textit{Static No-Scattering} model, the slope decreases until $250$ Myr and then slightly rises. 
At all times, most of the intrinsic stellar emission at UV and optical wavelengths comes from young and old stars \citep[stellar age $< 10$ and $> 10$ Myr, respectively;][]{Narayanan2018}. 
At $t=100$ Myr, dust is produced only by SNe and hence preferentially obscures only young stars in the star-forming region. 
Later, dust starts growing in the ISM due to metal accretion on dust grains and attenuates emission from old stars, increasing $A_V$ and flattening the slope until $t=250$ Myr. At higher inclination angles, older stars are likely to be obscured more by the dust along the line of sight, and hence, the slope further flattens until $t=500$ Myr. 
At later times, dust becomes optically thick to part of the UV emission from young stars, and only a small amount of UV emission from young stars leaks from the galaxy. Consequently, the UV observed spectrum is more contributed to by the emission from old stars. The intrinsic spectra of old stars become redder as the simulation time increases. As a result, the difference between the intrinsic UV spectra from young stars and the observed UV spectra from older stars leads to higher $A_{FUV}$ and a steeper attenuation curve \citep[see also Section 4.3 in][]{Narayanan2018}. 
After $t = 7$ Gyr, the intrinsic UV spectra are also increasingly dominated by older stars, resulting in lower $A_{FUV}$ and slightly flattening the slope.

\begin{figure*}[]
    \centering
    \includegraphics[width=0.95\textwidth]{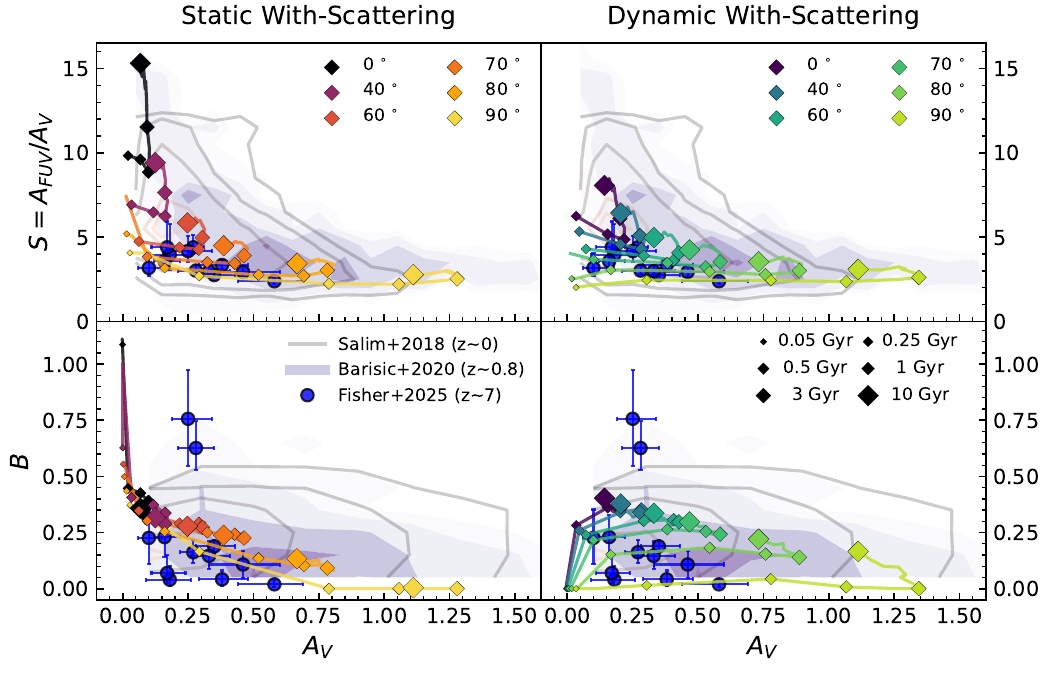}
    \caption{\textbf{Upper panels}: Relationships between the slope and $A_V$ for the \textit{Static With-Scattering} and \textit{Dynamic With-Scattering} models (left and right, respectively) at different inclination angles. Diamonds indicate simulation snapshots at $t = 0.05$, $0.25$, $0.5$, $1$, $3$, and $10$ Gyr, with sizes increasing with time. Solid lines connect diamonds at each inclination angle, representing evolutionary tracks. Gray and purple contours illustrate $25^\mathrm{th}$, $50^\mathrm{th}$, $75^\mathrm{th}$, $95^\mathrm{th}$, and $99^\mathrm{th}$ percentiles for the galaxies from the GSWLC observational catalog \citep{Salim2018} and the Large Early Galaxy Astrophysics Census survey \citep{Barisic2020}, respectively.    
    The blue symbols indicates the observational measurements made for high-redshift galaxies from the REBELS ALMA large program \citep{Fisher2025}.
    \textbf{Lower panels}: the relationships between the bump and $A_V$ for \textit{Static With-Scattering} and \textit{Dynamic With-Scattering} models (left and right panels, respectively). 
    }%
    \label{fig: the slope vs Av}
\end{figure*}

In the \textit{Dynamic No-Scattering} model, the slope increases up to $t=0.5$ Gyr, following the trend of the extinction slope. This is due to the formation of small grains ($a<0.1 \ \mathrm{\mu m}$) via the interplay between shattering and accretion in the ISM. Indeed, small grains absorb UV emission more significantly than larger dust grains, increasing $A_{FUV}$ and the slope. 
After $t=0.5$ Gyr, the small grains coagulate, forming larger grains, and as a consequence, the slope slightly decreases. Here, silicate grains are responsible for the attenuation at UV wavelengths and the change in the slope, as discussed in the evolution of the extinction curve in Section~\ref{sec: Evolution of the averaged extinction curve}.
Before $t=1$ Gyr, the evolution of the slope follows the trend of the average extinction curve, as shown by the gray thick line in Fig.~\ref{fig: the time evolution of slopes}.
Later, the grain size distribution becomes relatively stable over time, and the post-processed assigned ratio of silicate to carbonaceous grains shows little variation (see the bottom panel of Fig.~\ref{fig: Global Evolution of the galactic properties}). Consequently, the time evolution of the \textit{Dynamic No-Scattering} model closely resembles that of the \textit{Static No-Scattering}. As in the \textit{Static No-Scattering} model, the slope in the \textit{Dynamic No-Scattering} model decreases with inclination angle.

Once scattering is included (lower panels of Fig.~\ref{fig: the time evolution of slopes}), the time evolution of the slope deviates significantly from that of the extinction curve, with the extent of deviation depending on the inclination angle. The slopes for both \textit{Dynamic With-Scattering} and \textit{Static With-Scattering} models are significantly higher compared to the \textit{No-Scattering} models, especially at lower inclination angles. 
When the dust column density along the line of sight is low, optical photons can escape more easily from the galactic plane along the line of sight thanks to scattering, effectively reducing attenuation at optical wavelengths ($A_V$). However, the dust opacity at UV wavelengths is much higher than that at optical wavelengths, meaning that UV photons are less likely to escape even for the same dust column density independently from accounting for scattering or not. Thus, the attenuation at optical wavelengths is effectively reduced by scattering, resulting in an overall steepening of the attenuation curves. 
The effect of scattering is maximal when line-of-sight optical depth at optical wavelengths is low ($A_V<1$), and hence, the slope becomes steeper at lower inclination angles.
In fact, the attenuation curves with scattering are almost identical to those without scattering at the inclination angle of $90$ $^\circ$.
In addition, we also note that, as UV photons have stronger forward scattering by dust, they are unlikely to change their path and escape from the galactic disk \citep{Draine2003_Scattering}. 

In the \textit{Static With-Scattering} model, the slope becomes flatter up to $t=1$ Gyr, since the effectiveness of scattering at optical wavelengths is reduced as the optical depth along the line of sight increases due to dust formation via metal accretion. After $t=1$ Gyr, the slope becomes significantly steeper over time. Dust formation shifts from the center to the outskirts of the galaxy over time (see Fig. \ref{fig: Band maps at i00}) as the total dust mass monotonically increases. Consequently, the enhancement of dust obscuration of stars within the galaxy increases the possibility of scattering in the galactic plane, steepening the slope over time. This result suggests that the effect of scattering on attenuation curves is not solely determined by the dust column density along the line of sight. Rather, the dust column density perpendicular to the line of sight also plays a significant role in shaping the attenuation curve (we discuss the three-dimensional (3D) effect of scattering in Section~\ref{Sec: understanding three dimensional effects of scattering with a toy model} and \ref{Subsec: Physical mechanisms shaping the slope and UV bump of the attenuation curve} thoroughly).

The time evolutionary trend of the \textit{Dynamic With-Scattering} model is similar to that of the \textit{Static With-Scattering} model. However, the slope for the \textit{Dynamic With-Scattering} model starts with lower values at higher inclination angles due to the lack of small grains at $t < 0.5$ Gyr and the suppression of scattering processes. Furthermore, overall, the slope for the \textit{Dynamic With-Scattering} model is flatter compared to that for the \textit{Static With-Scattering} model. 
This is because large silicate dust grains ($a > 0.1 \ \mathrm{\mu m}$) are more populated in the \textit{Static With-Scattering} model \citep{Draine&Li2007}. Large silicate grains have a high albedo and are likely to scatter at optical wavelengths ($V$-band), resulting in the steeper attenuation curve for the \textit{Static With-Scattering} model.
These effects demonstrate the importance of the dust grain properties in shaping the attenuation curves.

In brief, in our simulations, the evolutionary trend of the slope for the \textit{Dynamic With-Scattering} model differs from that of the underlying extinction curve due to the interplay of multiple effects.
At low inclination angles, the attenuation curves are strongly shaped by scattering, which is sensitive to changes in the star–dust geometry.
At higher inclination angles, the effect of the scattering is suppressed, and thus, the attenuation curves are dominated by the variations in the star-dust geometry and grain size distribution.
Importantly, changes in grain size distribution affect not only the underlying extinction curve but also scattering processes, both of which influence the slope of the resulting attenuation curves. Appendix~\ref{app: Evolution of dust attenuation} further discusses the impacts of star-dust geometry, scattering, and dust properties on $A_V$ and $A_{FUV}$.

\subsection{Time evolution of the bump at various inclination angles}\label{sec: Time evolution of 2175 bump at various inclination angles}

We examine the main factors driving the time evolution of the 2175 \AA{} bump of the global attenuation curve in the MW-like galaxy simulation by comparing it with that of the extinction curve. 
Figure~\ref{fig: the time evolution of bump} shows the time evolution of the bump of the global attenuation curve of the MW-like galaxy simulation at various inclination angles. 

In the \textit{Static No-Scattering} model, the 2175 \AA{} bump strength is initially as high as in the extinction curve from $t=0.05$ to $0.1$ Gyr. Thereafter, it gradually decreases until $t=3$ Gyr and then remains nearly constant up to $t=10$ Gyr. 
Initially, dust remains optically thin to stellar emission at $\lambda = 2175$ \AA{}, meaning that most stars contribute to the observed emission with minimal differential attenuation across different lines of sight. As a result, the shape of the attenuation curve closely follows that of the intrinsic extinction curve, including the 2175 \AA{} bump. However, after $t=0.1$ Gyr, as dust column density increases, the UV optical depth exceeds unity ($A_{2175 \, \text{\AA{}}} > 1.0$) along some sight lines. This leads to a stronger bias toward emission from unobscured stars in the observed UV light \citep{Narayanan2018}, causing the observed attenuation curve to deviate from the intrinsic extinction curve and reducing the apparent strength of the 2175 \AA{} bump.
Consequently, the bump strength becomes weaker over time. 
At higher inclination angles, the area where dust is optically thick to the stellar emission at 2175 \AA{} increases, weakening the bump strength more.

In the \textit{Dynamic No-Scattering} model, the 2175 \AA{} bump strength gradually increases until $t=0.5$ Gyr due to the formation of small carbonaceous grains in the ISM via shattering and accretion. Here, small carbonaceous dust grains, including PAH and graphite, are responsible for the change in the bump strength as discussed in Section~\ref{sec: Evolution of the averaged extinction curve}. After $t=1.0$ Gyr, the bump strength becomes weaker because small grains are coagulated, and larger grains form. After $3$ Gyr, very small carbonaceous grains such as PAH ($a<0.0012$ $\mathrm{\mu m}$) begin to form through shattering, strengthening the 2175 \AA{} bump again. 
Similarly to the \textit{Static No-Scattering} model, the bump strength weakens at higher inclination angles, but the bump strength increases over time after $t=3.0$ Gyr at higher inclination angles. This is because the observed stellar emission at higher inclination angles is more affected by the dust grains in the outer disk of the galaxy, where small grains are more produced by shattering \citep{Matsumoto2024}. 

When scattering is included (lower panels of Fig.~\ref{fig: the time evolution of bump}), the 2175 \AA{} bump strength increases in both the \textit{Dynamic With-Scattering} and \textit{Static With-Scattering} models. 
The $2175$ \AA{} attenuation feature is primarily caused by small grains with near-zero albedo, including PAHs, whereas the surrounding wavelengths are attenuated by somewhat larger grains with non-zero albedo (see also Appendix \ref{app: Global evolution of albedo of the MW galaxy simulation}). As a result, scattering is more efficient at wavelengths near $2175$ \AA{} than at the feature itself. This leads to enhanced scattering out of the line of sight in the adjacent continuum, suppressing the surrounding attenuation and increasing the contrast of the $2175$ \AA{} bump in the attenuation curve, making it appear more pronounced.

The evolutionary trend of the $2175$ \AA{} bump strength for the \textit{Dynamic With-Scattering} model follows that of the extinction curve, even when scattering is taken into account. 
Therefore, we conclude that, while the $2175$ \AA{} bump strength is influenced by scattering and the star-dust geometry, the evolutionary trend of the $2175$ \AA{} bump strength still traces the change in the grain size distribution.
Remarkably, the attenuation curves in the \textit{Dynamic With-Scattering} dust model show a shallower or even absent $2175$ \AA{} bump in the galaxy at the early evolutionary phase of the galaxy ($t<250$ Myr), as the interplay of shattering and accretion has not yet fully been activated in the ISM.
Therefore, the $2175$ \AA{} bump strength can serve as a good proxy for the mass fraction of small carbonaceous grains relative to the total dust mass in galaxies, at a given inclination angle.

\subsection{$A_V$ vs slope and bump} \label{Subsec: Effect of scattering: A_V vs slope and bump}
The upper panels of Fig.~\ref{fig: the slope vs Av} show the relationship between the slope and $A_V$ at different inclination angles for the \textit{Static With-Scattering} and \textit{Dynamic With-Scattering} models.
The gray and purple contours in each panel of Fig.~\ref{fig: the slope vs Av} show the relationships between slope and $A_V$ obtained from observations of local galaxies with stellar masses of $M_*=10^9-10^{11}$ $\Msun$ from the GALEX SDSS-WISE Legacy Catalog (GSWLC; \citealt{Salim2018}) and star-forming galaxies at relatively higher redshifts of $z\sim0.8$ from the Large Early Galaxy Astrophysics Census survey \citep[LEGA-C survey;][]{Barisic2020}, respectively. 

The relationships between the slope and $A_V$ for both \textit{Static With-Scattering} and \textit{Dynamic With-Scattering} models show a similar trend to those of the observations. Overall, the slope becomes steeper at lower $A_V$. Indeed, as the line-of-sight optical depth at optical wavelengths decreases, optical photons easily escape along the line of sight from the galactic plane, decreasing $A_V$ and steepening the slope, especially at lower inclination angles.
These results confirm that the decreasing trend of the slope against $A_V$ is explained by the effect of scattering \citep[][we further discuss in Section \ref{Subsec: Physical mechanisms shaping the slope and UV bump of the attenuation curve}]{Chevallard2013} and is consistent with the observational results from \citet{Salim2018} as well as the ones from simulations in a static dust model framework and at lower resolution presented by \cite{Trayford2020} and \cite{Sommovigo2025}. This relationship is seen not only in nearby galaxies \citep{Salim2018, Leja2017} but also in higher redshift galaxies \citep{Barisic2020,Boquien2022}. 
Moreover, compared to the \textit{Static With-Scattering} model, the \textit{Dynamic With-Scattering} model produces flatter attenuation curves at lower $A_V$, as observed in high-redshift galaxies \citep{Fisher2025}, since small grains are absent in the early evolutionary stages of galaxies.
\begin{figure*}[htbp]
    \centering
    \includegraphics[width=0.95\textwidth]{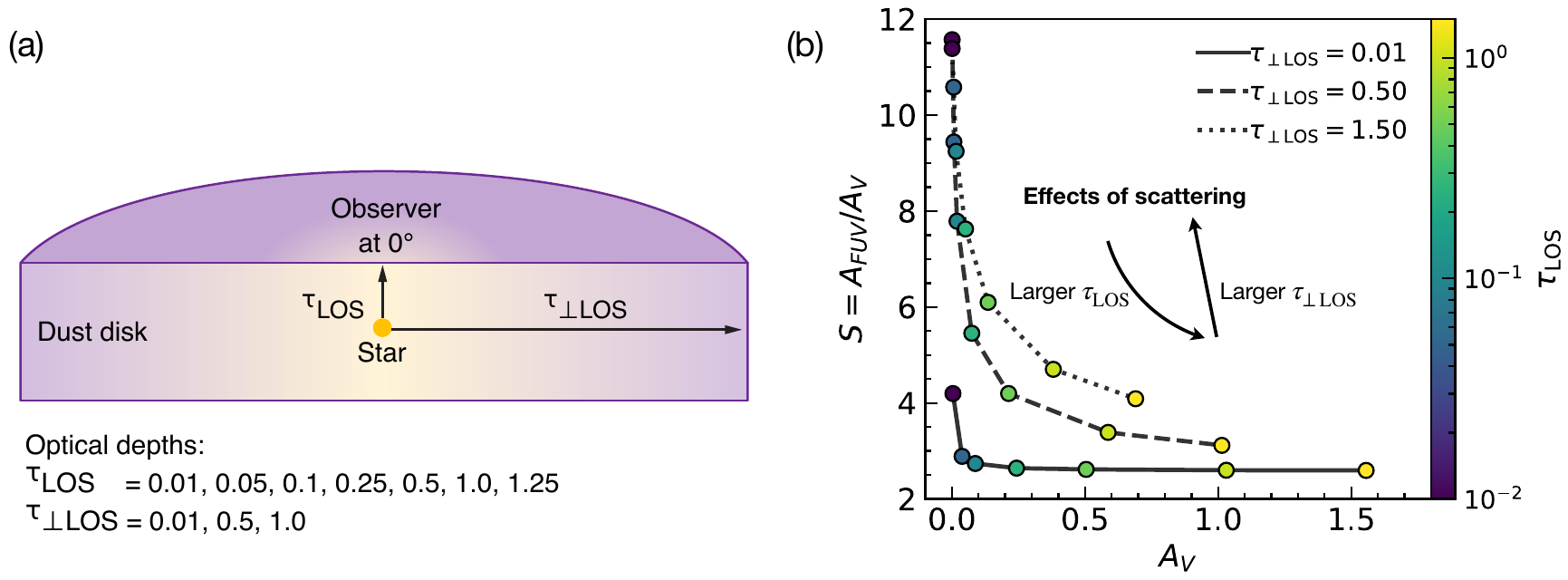}
    \caption{\textbf{(a)} Schematic picture of a toy model with a single star embedded in a dust disk. $\tau_\mathrm{LOS}$ and $\tau_\mathrm{\perp \, LOS}$ are optical depths at $V$-band ($\lambda=5500$ \AA{}) along the line of sight and perpendicular to the line of sight. The two parameters that serve to vary the dust geometry. The resulting attenuation curves from this model are influenced solely by scattering processes, which vary in response to changes in the star–dust geometry. \textbf{(b)} Relationship between $A_V$ and slope based on this simple toy model. Different line styles represent different $\tau_\mathrm{\perp LOS}$. 
    }%
    \label{fig: toy model slope vs Av}
\end{figure*}

The lower panels of Fig.~\ref{fig: the slope vs Av} show the relationship between the 2175 \AA{} bump strength and $A_V$ at different inclination angles for the \textit{Static With-Scattering} and \textit{Dynamic With-Scattering} models. 
A larger fraction of the dust becomes optically thick to stellar emission at UV wavelengths at higher $A_V$, resulting in a stronger bias toward emission from unobscured stars in the observed UV light \citep{Narayanan2018}, thereby reducing the apparent bump strength.
In the \textit{Dynamic With-Scattering} model, the bump strength is initially low and increases over time as small dust grains form within the galaxy. However, its peak strength diminishes with increasing $A_V$ or inclination angle, similar to the behavior seen in the \textit{Static With-Scattering} model.
While the \textit{Static With-Scattering} model exhibits a tight correlation between the bump strength and $A_V$, the \textit{Dynamic With-Scattering} model predicts a broader distribution of bump strengths and $A_V$, varying with time and inclination angle, and encompassing the observational results of \citet{Barisic2020}. In contrast, no clear trend is observed in the data from \citet{Salim2019}, likely due to the $2715$ \AA{} bump falling between the GALEX/FUV and NUV-bands in photometric observations of local galaxies, which complicates its accurate estimation \citep[see Appendix A.1 of][]{Salim2018}.
In the \textit{Dynamic With-Scattering} model, young galaxies with low $A_V$ exhibit shallower or absent bumps in their attenuation curves, regardless of the inclination angle. This finding is consistent with recent observations of galaxies at redshifts $z>6$ \citep{Fisher2025}, which also show the absence or shallowness of these bumps in galaxies with low $A_V$. This could suggest a lack of small grains or a small amount of them in the early evolutionary stages of galaxies.



\section{Discussion}\label{Discussion}
\subsection{Understanding the 3D effect of scattering with a toy model}
\label{Sec: understanding three dimensional effects of scattering with a toy model}

In Section~\ref{Subsec: Effect of scattering: A_V vs slope and bump}, we show that the attenuation curve becomes steeper over time in our MW-like galaxy simulation due to the increasing possibility of scattering as the dust mass increases.
This behavior reflects the 3D radiative effect of scattering, which depends not only on the optical depth along the line of sight but also on the optical depth perpendicular to it.
To demonstrate the effect of scattering on the slope of the attenuation curve, we consider a toy model with a single star embedded in a dust disk as shown in Fig.~\ref{fig: toy model slope vs Av}(a). Here, dust properties are taken from \citet{Draine&Li2007}, and the geometry of the dust disk is parameterized by the optical depths at the $V$-band ($5500$ \AA{}) along the line of sight and perpendicular to the line of sight ($\tau_\mathrm{LOS}$ and $\tau_\mathrm{\perp LOS}$, respectively). 
We then conduct 3D radiative transfer calculations using SKIRT for this toy model to calculate the attenuation curve for the star. 
In this model, only scattering influences the attenuation curves.


Figure~\ref{fig: toy model slope vs Av}(b) shows the relationship between $A_V$ and the slope of the attenuation curves based on the radiative transfer calculations for the simple toy model. 
The slope decreases with increasing $A_V$ because, at lower $\tau_\mathrm{LOS}$, scattered photons from the central star can more easily escape along the line of sight. This causes the observed $A_V$ to be lower than the actual optical depth $\tau_\mathrm{LOS}$, resulting in a steeper attenuation curve.
This effect is further enhanced at higher optical depths perpendicular to the line of sight ($\tau_{\perp \mathrm{LOS}}$), where scattering is more likely to happen in the dusty disk. As a result, the observed $A_V$ becomes lower than the actual $\tau_\mathrm{LOS}$, further steepening the attenuation curve.
Therefore, the relation between $A_V$ and slope and the associated scatter arises from the configuration of dust optical depths along and perpendicular to the line of sight.
These two optical depths are ultimately governed by the inclination angle of galaxies and the star-dust geometry.

\subsection{Physical mechanisms shaping attenuation curve} \label{Subsec: Physical mechanisms shaping the slope and UV bump of the attenuation curve}


\begin{figure*}[htbp]
    \centering
    \includegraphics[width=1.0\textwidth]{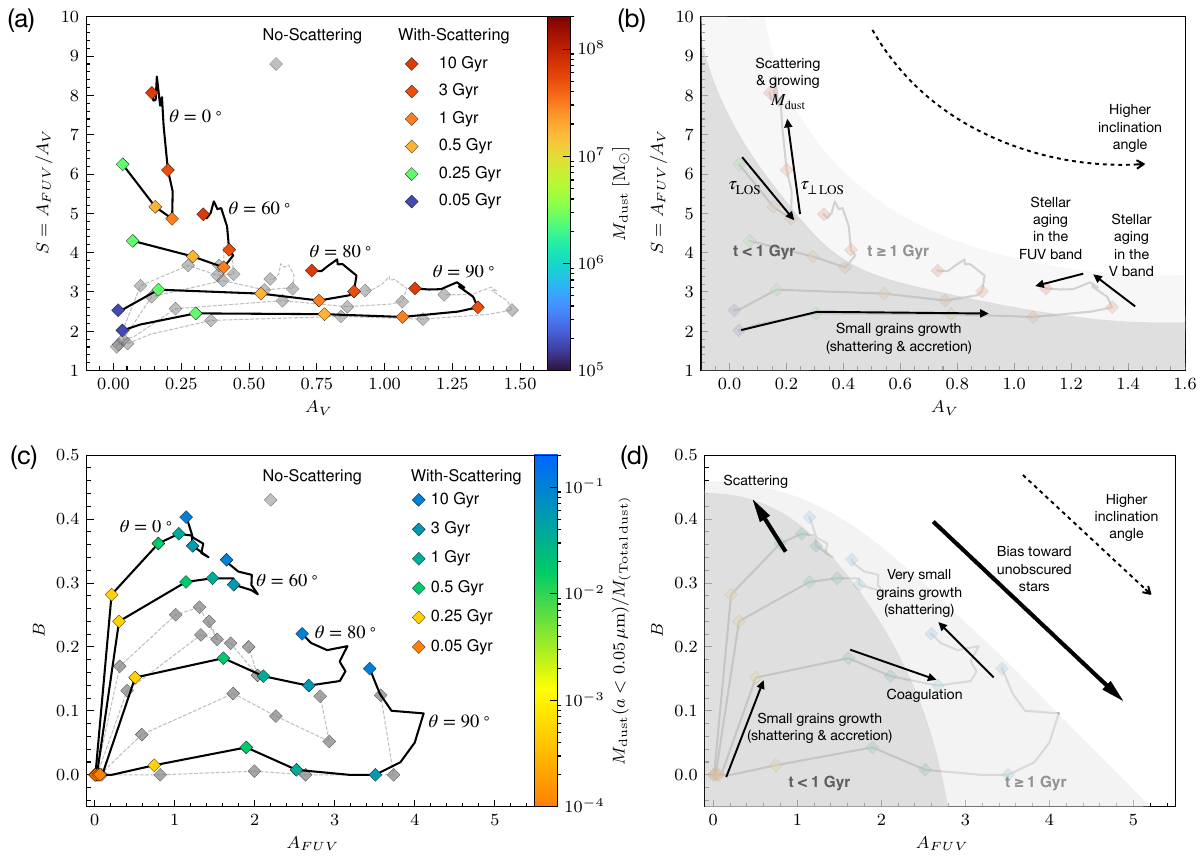}
    \caption{\textbf{(a)} Relation between the slope and $A_V$. Colored diamonds represent the \textit{Dynamic With-Scattering} model, while gray diamonds shows the \textit{Dynamic No-Scattering} model. The color indicates the total dust mass within the galaxy for the \textit{Dynamic No-Scattering} model. Diamonds correspond to simulation snapshots at $t = 0.05$, $0.25$, $0.5$, $1$, $3$, and $10$ Gyr. For the \textit{Dynamic With-Scattering} and \textit{No-Scattering} models, solid and dashed lines connect the diamonds, respectively, representing the evolutionary tracks at each inclination angle. 
    \textbf{(b)} Physical explanation of the relation between the slope and $A_V$. Solid arrows indicate the effects shaping the evolution of the attenuation curve slope, while the dashed arrow highlights the dependence of the slope on the inclination angle. The dark gray-shaded region corresponds to the early phase of our MW-like galaxy ($t < 1$ Gyr), whereas the light gray-shaded region marks the later phase ($t \geq 1$ Gyr).
    \textbf{(c)} Relation between the $2175$ \AA{} bump strength and $A_{FUV}$. The color indicates the mass fraction of small grains ($a < 0.05$ $\mathrm{\mu m}$) relative to the total dust mass for the \textit{Dynamic No-Scattering} model.
    \textbf{(d)} Physical explanation of the relation between the $2175$ \AA{} bump strength and $A_{FUV}$. Thin solid arrows indicate the effects shaping the evolution of the $2175$ \AA{} bump, while thick solid arrows represent effects that systematically change the trend across inclination angles or $A_{FUV}$. The dashed arrow highlights the dependence of the slope on the inclination angle.
    }
    \label{fig: summary of effects of the attenuation curves}
\end{figure*}

Figure~\ref{fig: summary of effects of the attenuation curves}(a) shows the relation between the slope and $A_V$ for the \textit{Dynamic With-Scattering} and \textit{Dynamic No-Scattering} models.
The relation for the \textit{Dynamic No-Scattering} exhibits a flat trend, while the \textit{Dynamic With-Scattering} shows a decreasing trend of the slope with increasing $A_V$ due to scattering, as discussed in Section~\ref{Subsec: Effect of scattering: A_V vs slope and bump}.
For further discussion, the key processes influencing the slope are schematically illustrated in Fig.~\ref{fig: summary of effects of the attenuation curves}(b). 
At lower inclination angles, optical photons escape along the line of sight via scattering, which reduces $A_V$ and steepens the attenuation curve slope. 
Up to $t=1$ Gyr the slope decreases with increasing $A_V$,driven by the growing optical depth along the line of sight ($\tau_\mathrm{LOS}$). 
After $t=1$ Gyr, the trend reverses: the slope increases with decreasing $A_V$. This is because the optical depth \textit{perpendicular} to the line of sight increases ($\tau_\mathrm{\perp \, LOS}$), enhancing the probability of scattering within the galactic plane. 
We confirm that the decrease in $A_{V}$ during this later phase is due to scattering effects rather than a decline in dust column density along the line of sight over time (see Appendix~\ref{app: Evolution of dust attenuation}).
Thus, changes in the star–dust geometry modulate the efficiency and directionality of scattering, producing time-dependent variations in both the slope and $A_V$, as also demonstrated in Section~\ref{Sec: understanding three dimensional effects of scattering with a toy model} with a toy model.
As a result, at lower inclination angles, galaxies with higher dust masses at a given $A_V$ tend to exhibit steeper slopes in our simulation.
We also note that during this period large grains ($a>0.1$ $\mathrm{\mu m}$) begin to form through coagulation, enhancing the probability of scattering in the $V$-band.
However, this trend can be significantly degenerate with that produced by large variations in the star–dust geometry \citep[][]{Narayanan2018, Trayford2020, Sommovigo2025}.

At higher inclination angles, scattering becomes less effective, and the attenuation curve slope is instead shaped by changes in the grain size distribution and star-dust geometry-- particularly through the fraction of obscured young stars \citep{Narayanan2018}, on the slope becomes important.
Up to $t = 3$ Gyr, the slope steepens slightly with increasing $A_V$, as young stars are preferentially obscured by dust and small grains form through shattering within the galaxies. During this phase, the observed $V$-band flux is dominated by older stars, while the intrinsic flux includes contributions from both young and old stars.
Also, at earlier times ($t \leq 0.5$ Gyr), the slope remains relatively flat due to the lack of small grains in the galaxy, which reduces $A_{FUV}$.
After $t = 3$ Gyr, the star formation rate declines significantly, and the intrinsic $V$ band emission becomes increasingly dominated by older stars. As a result, the intrinsic and observed light originate from similar stellar populations, reducing the difference between them, which lowers $A_V$ and steepens the slope.
After $t = 7$ Gyr, the intrinsic $FUV$ band emission is also increasingly dominated by older stars, resulting in lower $A_{FUV}$ and slightly flattening the slope.
Overall, the relative importance of scattering, star–dust geometry, and grain size distribution varies non-trivially depending on $A_V$ and inclination angles.



Figure~\ref{fig: summary of effects of the attenuation curves}(c) shows the relationship between the $2175$ \AA{} bump strength and $A_{FUV}$ for the \textit{Dynamic With-Scattering} and \textit{Dynamic No-Scattering} models. We use $A_{FUV}$ instead of $A_V$ because it is less affected by scattering. 
The $2175$ \AA{} bump strength in the \textit{Dynamic With-Scattering} model is slightly higher than in the \textit{Dynamic No-Scattering} model. This is because scattering is more efficient at wavelengths adjacent to $2175$ \AA{} than at the bump itself (see Appendix \ref{app: Global evolution of albedo of the MW galaxy simulation}). Enhanced scattering reduces the attenuation in the surrounding continuum, and increases the contrast of the bump in the attenuation curve.
For further discussion, the key processes affecting the bump strength are schematically illustrated in Fig.~\ref{fig: summary of effects of the attenuation curves}(d).
At low inclination angles, the bump strengths dramatically increase over time up to $t=0.5$ Gyr, as small grains ($a < 0.05$ $\mathrm{\mu m}$) form in the galaxy through the interplay between accretion and shattering.
After $t=0.5$ Gyr, small grain formation becomes less efficient, since some small grains start being coagulated into larger grains, causing the bump strength to nearly plateau.
After $t=1.0$ Gyr, coagulation slightly weakens the bump strength with time.
At the same time, as dust column density increases with time, the UV optical depth exceeds unity ($A_{2175 \, \text{\AA{}}} > 1.0$) along some sight lines. This leads to a stronger bias toward emission from unobscured stars in the observed UV light, which further reduces the apparent bump strength \citep{Narayanan2018}.
After $t=5$ Gyr, very small carbonaceous grains such as PAH ($a<0.0012$ $\mathrm{\mu m}$) begin to form through shattering, leading to a renewed increase in bump strength.

At higher inclination angles, the bias toward emission from unobscured stars becomes more pronounced, systematically reducing the observed bump strength.
Early on (up to $t = 0.5$ Gyr), the bump strength shows a slight increase due to the formation of small carbonaceous grains. Afterward, as the dust column density grows, this leads to higher UV optical depths and an enhanced bias toward unobscured stellar emission.
After $t = 3$ Gyr, the bump strength significantly increases again. This is because small carbonaceous grains preferentially form in the diffuse ISM, which is more likely to be along the line of sight at higher inclination angles.
After $t = 7$ Gyr, the intrinsic FUV emission becomes increasingly dominated by older stars, leading to a reduction in $A_{FUV}$.
Together, these effects highlight the critical roles of grain size evolution and star–dust geometry in shaping the strength and visibility of the $2175$ \AA{} bump over time.

\subsection{Comparison with theoretical studies based on
large-volume cosmological simulations}\label{sec: Comparison with theoretical studies based on large-volume cosmological simulations}
In this study, we demonstrated a strong relation between $A_V$ and the slope of the attenuation curves. The scatter in this relation is attributed to variations in inclination angles and in the configuration of dust optical depths around stars. (see Fig.~\ref{fig: the slope vs Av} and Fig.~\ref{fig: summary of effects of the attenuation curves}(a)). 
However, we handle only a MW-like galaxy simulation and lack the variation of star-dust geometry.
Previous studies based on large-volume cosmological simulations offer an advantage in capturing the variation of star-dust geometry, although they typically assume a fixed dust-to-metal ratio and constant dust properties based on dust models \citep[e.g.,][]{Weingartner2001, Zubko2004, Draine&Li2007}, and are limited by lower spatial resolution.
\citet{Trayford2020} identified a correlation between $A_V$ and the UV slope in EAGLE galaxies across redshifts $z=0$–$2$, with variations attributed to galaxy inclination. Specifically, galaxies viewed at higher inclination angles tend to exhibit flatter attenuation curves. This trend is consistent with our findings, as well as with results from local galaxies in TNG simulations \citep{Sommovigo2025} and high-redshift galaxies in the FirstLight and MUFASA simulations \citep[][respectively]{Mushtaq2023, Narayanan2018}.

Furthermore, the importance of the star-geometry is discussed by examining the factors to change $A_V$. \citet{Trayford2020} found that $A_V$ is predominantly influenced by dust surface densities of galaxies. 
\citet{Mushtaq2023} found that galaxies with higher stellar masses contain a larger amount of dust. Concurrently, \citet{Sommovigo2025} revealed that galaxies with higher $\Sigma_\mathrm{SFR}$ (i.e., compact star-forming galaxies) tend to show higher $A_V$. These results suggest that $A_V$ is mainly driven by the star-dust geometries of galaxies. As a result, the star-dust geometry changes the slope of the attenuation curves according to the relation between $A_V$ and slope.
\citet{Narayanan2018} provided physical insights into variations of the slope of attenuation curves. The relative fraction of dust-obscured young stars compared to non-obscured old stars changes the shape of observed spectra of galaxies at UV wavelengths: galaxies with heavily obscured young stars exhibit steeper curves, as the observed UV light is dominated by older stellar populations. 
In our simulation, we confirm that changes in the dust obscuration fraction of young stars affect the slope of the attenuation curves in the \textit{Static No-Scattering} model (see the upper left panel in Fig.~\ref{fig: the time evolution of slopes}). This effect, however, becomes less pronounced when scattering is included, as scattering significantly changes the slope of the attenuation curve.
We note that this lower responsiveness of the pure star–dust geometry effect might be due to the limited variation in star–dust geometry in our simulations. Therefore, a more comprehensive investigation using a large suite of galaxy simulations is required to fully assess the impact of star–dust geometry on attenuation curves along with dust evolution.

Moreover, we found that a decreasing trend of the $2175$ \AA{} bump strength with $A_V$ in the \textit{Static With-Scattering} model (see the lower left panel in Fig.~\ref{fig: the slope vs Av}) is consistent with \citet{Narayanan2018} and \citet{Sommovigo2025}.
\citet{Narayanan2018} found a correlation between the fraction of unobscured young stars in galaxies and the bump strengths, suggesting the importance of the effect of the star-dust geometry. Specifically, \citet{Seon2016} argued that an inhomogeneous distribution of optically thick clumps obscuring stars weakens the $2175$ \AA{} bump.
Additionally, \citet{Narayanan2018} indicated that the bump strength slightly decreases when scattering is less effective in galaxies and concluded that scattering is a minor effect that influences the bump strength. These findings agree with our results.

\subsection{Comparison with observational studies}
\label{Subsec: Comparison with observational studies}
The relation between $A_V$ and the slope of the attenuation curve is also reported in observational studies \citep[e.g.,][]{Salmon2016, Leja2017, Salim2018, Boquien2022}.
We argued that the scatter in the relation arises from variations in galaxy inclination: the attenuation curves at higher inclination angles tend to be flatter and exhibit less pronounced $2175$ \AA{} bumps. 
Observationally, \citet{Battisti2017b} found flatter curves and weaker bumps at higher inclination angles in local star-forming galaxies, consistent with our results \citep[see also][]{Wild2011}. This trend is also confirmed at higher redshifts \citep[$z \sim 0.8$;][]{Barisic2020}.

In the lower right panel of Fig.~\ref{fig: the slope vs Av}, the comparison of the relation between $A_V$ and $2175$ \AA{} bump with observations \citep{Battisti2017b} shows that the \textit{Dynamic With-Scattering} model reproduces the observed decreasing trend of the maximum bump strength with $A_V$, which is also found in high redshift observed galaxies \citep{Markov2025}. We attributed this trend to a stronger bias toward unobscured stellar emission in the observed UV light at higher UV optical depths \citep{Narayanan2018}. We also argued that the bump strength increases according to the formation of small grains within galaxies via the interplay between shattering and accretion, reproducing the observed distribution of the bump strengths and $A_V$ \citep{Barisic2020}.

Recent JWST observations have revealed the characteristics of the attenuation curves in galaxies at high redshifts ($z>4$). \citet{Markov2024} show the average attenuation curves steepening with cosmic time, indicating that galaxies at higher redshifts of $z=4.5$-$11.5$ exhibit flatter attenuation curves with less pronounced $2175$ \AA{} bumps. Additionally, \citet{Fisher2025} exhibits flat attenuation curves from $12$ Lyman-break galaxies at $z=6.5-7.7$ with weak $2175$ \AA{} bump. Notably, high-redshift galaxies tend to exhibit flatter curves even at low $A_V$ compared to nearby galaxies \citep[see also][]{Markov2025}. 
\citet{Markov2024} suggested that the flat attenuation curves observed at high redshifts are driven by large grains produced by supernovae \citep{McKinney2025}, which have low UV opacity. This scenario also provides an explanation for the weaker $2175$ \AA{} bump seen in the average attenuation curves at high redshifts.
In the \textit{Dynamic With-Scattering} model, we find that the attenuation curves exhibit shallower slopes and weaker $2175$ \AA{} bumps during the early stages of galaxy evolution (see the lower right panels in Fig.~\ref{fig: the time evolution of slopes} and \ref{fig: the time evolution of bump}).
In our simulations, attenuation curves at high inclination angles tend to be flatter during the early epoch ($t < 1$ Gyr), due to the lack of small grains.
At low inclination angles, attenuation curves are more strongly influenced by scattering and thus appear steeper. In the early stages of galaxy evolution, however, the effect of scattering is reduced.
This is because the dust distribution remains compact in three dimensions at early times ($t < 1$ Gyr) -- unlike the extended dusty disk that forms later (see Fig.~\ref{fig: Band maps at i00}) -- resulting in weaker scattering and correspondingly flatter attenuation curves.

On the other hand, our \textit{Dynamic With-Scattering} model initially shows a weak $2175$ \AA{} bump, which becomes more prominent on a timescale of $\sim250$ Myr as small grains ($< 0.05$ $\mathrm{\mu m}$) are produced through the interplay between accretion and shattering, consistent with the scenario proposed by \citet{Markov2024}.
\citet{Ormerod2025} reported the detection of a $2175$ \AA{} bump in a Lyman-break galaxy at $z = 7.1$, indicating the rapid formation of small grains. 
They proposed the presence of an older stellar population ($\sim$252 Myr), suggesting that small grains could form through shattering processes following initial star formation. In contrast, \citet{Witstok2023} identified a strong bump in a galaxy at $z = 6.7$ without signs of an old stellar population, implying that small grains may form shortly after the onset of star formation, potentially through mechanisms such as carbonaceous grain production in Wolf–Rayet stars \citep{Lau2023,Lau2024}.

\subsection{Limitations of our simulation}
\label{subsec: Implication and limitation of our simulation}

First, we use an isolated galaxy simulation with an initial stellar disk and bulge, and a dust-free gas disk. Without cosmological effects like mergers and with steady IGM inflow, the galaxy follows a moderate star formation history and weak stellar feedback.
Thus, our model does not fully capture the diversity of star-dust geometry seen across galaxy populations (see Section~\ref{sec: Comparison with theoretical studies based on large-volume cosmological simulations}).


Second, our simulation does not have sufficient spatial resolution to resolve the star formation processes embedded in individual molecular clouds (MCs), which require sub-pc resolution. \citet{DiMascia2025} discussed the significant impact of dust evolution in MCs on the attenuation curves of these regions. Additionally, many observations report that the optical reddening, $E(B-V)$, for nebular emission lines from star-forming regions is larger than that for optical stellar continuum emission \citep[e.g.,][]{Calzetti1994, Calzetti2000, Buat2018, Reddy2015, Reddy2020, Shivaei2020}, implying the importance of modeling the steep attenuation curve in birth clouds \citep{CharlotFall2000}. 
On the other hand, \citet{Trayford2020} demonstrated that modeling the steep attenuation curve in birth clouds does not significantly affect galaxy attenuation curves in EAGLE simulations.

Finally, our dust evolution model also contains some uncertainties. In our previous studies, \citet{Matsumoto2024} compared the dust properties from our simulations with observations of nearby galaxies and found that the model underestimates the abundance of very small grains such as PAHs ($a < 0.0013$ $\mathrm{\mu m}$). In contrast, \citet{vanderGiessen2024} reported that our model overestimates the amount of small grains with sizes $a < 0.015$ $\mathrm{\mu m}$. These findings suggest that the \citet{Draine&Li2007}-like grain size distribution for carbonaceous grains characterized by distinct peaks at both small and large sizes ($a = 0.0006$ and $0.4$ $\mathrm{\mu m}$, respectively) may be more appropriate. 
Furthermore, our post-processing dust decomposition recipe does not account for variations in dust accretion efficiency depending on the type of metal elements, nor for the depletion of key elements such as O, Si, Fe, and C within galaxies \citep[see][]{Choban2022, Dubois2024}.
\citet{Li2021} found that the graphite-to-silicate ratio of dust grains influences the strength of the 2175 \AA{} bump in the extinction curve of a MW-like galaxy simulation.
\citet{Dubois2024} reported that the accretion efficiency of carbonaceous grains is lower than that of silicate grains, implying that the emergence of the prominent 2175 \AA{} bump, which appears on a timescale of $\sim250$ Myr in our model, could be delayed in their model.
In addition, the extinction curve in our simulation at $t=10$ Gyr is steeper than that for the MW observation \citep[][see Fig.~\ref{fig: the evolution of the extinction curve}]{Pei1992}. This is because our model does not account for the depletion of key metal elements, potentially leading to an overestimation of the amount of silicate grains.
Therefore, future simulations should incorporate dust composition evolution for a more accurate treatment.


\section{Conclusions} \label{Conclusion}
In this paper, we have investigated the average extinction curve and global attenuation curves of a MW-like galaxy simulation and examined the physical mechanisms that shape the attenuation curves. To accomplish this, we performed post-processing dust radiative transfer calculations using the SKIRT code, based on the MW-like galaxy simulation \citep{Matsumoto2024}. 
Our model incorporates the evolution of grain size distributions across $30$ sizes from $3.0 \times 10^{-4}$ to $10 \ \mathrm{\mu m}$ and considers the hydrodynamic evolution of the galaxy, including star formation and the OSAKA feedback model. For modeling the dust composition, we decompose the dust grains into silicate and carbonaceous components, as well as PAHs, based on the abundance of Si and C and the mass fraction of dense gas in each gas particle in post-processing.
To disentangle the impacts of star-dust geometry, scattering, and grain size distribution on the attenuation curves, we have generated four different types of attenuation curves (\textit{Static No-Scattering}, \textit{Static With-Scattering}, \textit{Dynamic No-Scattering}, and \textit{Dynamic With-Scattering}). The details of the model prescription are explained in Section \ref{Sec: Evolution of the global attenuation curve}.

We summarize our results as follows:

\begin{enumerate}

    \item 
    The extinction curve evolves from flat to steep by $t = 0.5$ Gyr, gradually developing the $2175$ \AA{} bump, driven by the formation of small grains via shattering and accretion. It then flattens due to coagulation up to $t = 3$ Gyr. Later, the extinction curve slightly steepens again up to $t = 10$ Gyr, and the $2175$ \AA{} bump grows as PAHs form in the ISM (see Fig.~\ref{fig: the evolution of the extinction curve}).

    \item
    At $t = 0.1$ Gyr, the extinction curve is dominated by silicate grains across all wavelengths. However, at later times, carbonaceous grains become the dominant contributors to extinction at wavelengths longer than $2000$~\AA{} (see Fig.~\ref{fig: the evolution of the extinction curve of each component}).

    \item 
    At low inclination angles, the attenuation curve for the \textit{Dynamic With-Scattering} model evolves differently from the extinction curve because of scattering. 
    The attenuation curves slightly flatten over time up to $t = 1$ Gyr due to the increasing dust column density along the line of sight, and later, they become progressively steeper due to the interplay between scattering and extending dust disk geometry (see Fig.~\ref{fig: the time evolution of slopes}).

    \item 
    At high inclination angles, the attenuation curves starts with lower values at higher inclination angles due to the lack of small grains at $t < 0.5$ Gyr. At later times, the attenuation curves become slightly steeper due to differences in the contributions of young and old stars to the observed and intrinsic spectra and the formation of small grains.

    \item 
    Comparing attenuation curves between the \textit{Dynamic With-Scattering} and \textit{Static With-Scattering} models shows that the underlying extinction curve influences attenuation curves at high inclination angles and during the early epoch ($t<1$ Gyr). The attenuation curves for the \textit{Dynamic With-Scattering} is significantly flatter than \textit{Static With-Scattering} due to the lack of small grains.    
    At low inclination angles, variations in the grain size distribution, which affect the albedo, influence the slope primarily through scattering (see also Appendix~\ref{app: Global evolution of albedo of the MW galaxy simulation} and \ref{app: Evolution of dust attenuation}). These results highlight the importance of dust properties—both composition and grain size distribution—in determining the shape of attenuation curves.

    \item 
    The interplay between scattering and star-dust geometry plays a key role in shaping the relation between the slope and $A_V$, as seen in observations \citep[Fig.~\ref{fig: the slope vs Av};][]{Salim2018, Barisic2020}.
    The scatter of this relation arises due to the dust optical depth along and perpendicular to the line of sight, and thus reflects the inclination angle and star-dust geometry (see Fig.~\ref{fig: toy model slope vs Av}, \ref{fig: summary of effects of the attenuation curves}(a) and (b)). Changes in the grain size distribution and the star–dust geometry \citep{Narayanan2018} also play a key role in shaping slope of the attenuation curves at high inclination angles or high $A_V$ $(>0.5)$.
    
    \item 
    At lower inclination angles, the $2175$ \AA{} bump becomes stronger on a timescale of $\sim 250$ Myr due to the formation of small carbonaceous grains ($a < 0.05$ $\mu$m) through the interplay between shattering and accretion  (see Fig.~\ref{fig: the time evolution of bump}). After $t=1.0$ Gyr, the bump strength become lower as small grains are coagulated. At the same time, as dust column density along the line of sight increases with time, the bump strength becomes weaker due to the stronger bias toward emission from unobscured stars -- an effect governed by star-dust geometry.
    After $t=5$ Gyr, very small carbonaceous grains such as PAH ($a<0.0012$ $\mathrm{\mu m}$) begin to form through shattering, enhancing bump strength.
    At higher inclination angles, the bias toward emission from unobscured stars becomes more pronounced, further reducing the observed bump strength.

    \item 
    Variation in the bump strength is mainly caused by both the variation in the grain size distributions and star-dust geometry (see Fig.~\ref{fig: summary of effects of the attenuation curves}(c) and (d)).
    The maximum bump strength become systematically weaker at higher $A_{FUV}$ or higher inclination angles because of the bias toward emission from unobscured stars.


\end{enumerate}
These results provide physical insights into recent JWST observations, which reveal flatter attenuation curves and less prominent $2175$ \AA{} bumps in galaxies at high redshifts \citep[$z > 4$;][]{Markov2023,Markov2024,Markov2025,Fisher2025}. However, the systematic evolution of the attenuation curves could vary depending on the diversity of galaxies, especially star-dust geometry. Therefore, to better understand how dust evolution impacts the systematic evolution of attenuation curves, future studies using cosmological simulations that incorporate dust evolution across a wide range of galaxy types will be crucial.

\paragraph{Data Availability Statement}
The data used in this paper is publicly available\footnote{\url{https://github.com/Koseimatsu/MW_Galaxy_Properties_2025.git}}.

\begin{acknowledgements}
We thank Dr. Hirashita and Prof. Nakagawa for their valuable discussions.
We are grateful to Lucy Reading-Ikkanda for her invaluable help in creating and refining the schematic figures for this paper.
KM is a Ph.D. fellow of the Flemish Fund for Scientific Research (FWO-Vlaanderen) and acknowledges the financial support provided through Grant number 1169822N.
Numerical computations were performed on the Cray XC50 at the Center for Computational Astrophysics, National Astronomical Observatory of Japan, on the SQUID at the Cybermedia Center, Osaka University, as part of the HPCI system Research Project (hp220044, hp230089, hp240141), and on the Flatiron Institute’s research computing facilities (Rusty).
This work is supported in part by the MEXT/JSPS KAKENHI grant numbers 20H00180, 22K21349, 24H00002, 24H00241 (KN). 
KN acknowledges the travel support from the Kavli IPMU, World Premier Research Center Initiative (WPI), where a part of this work was conducted. 
\end{acknowledgements}

\bibliography{dust_paper}    
\begin{appendix} 
\section{Global properties of the MW-like galaxy simulation} \label{app: Global properties of the MW galaxy simulation}
\begin{figure}[]
    \centering
    \includegraphics[width=0.45\textwidth]{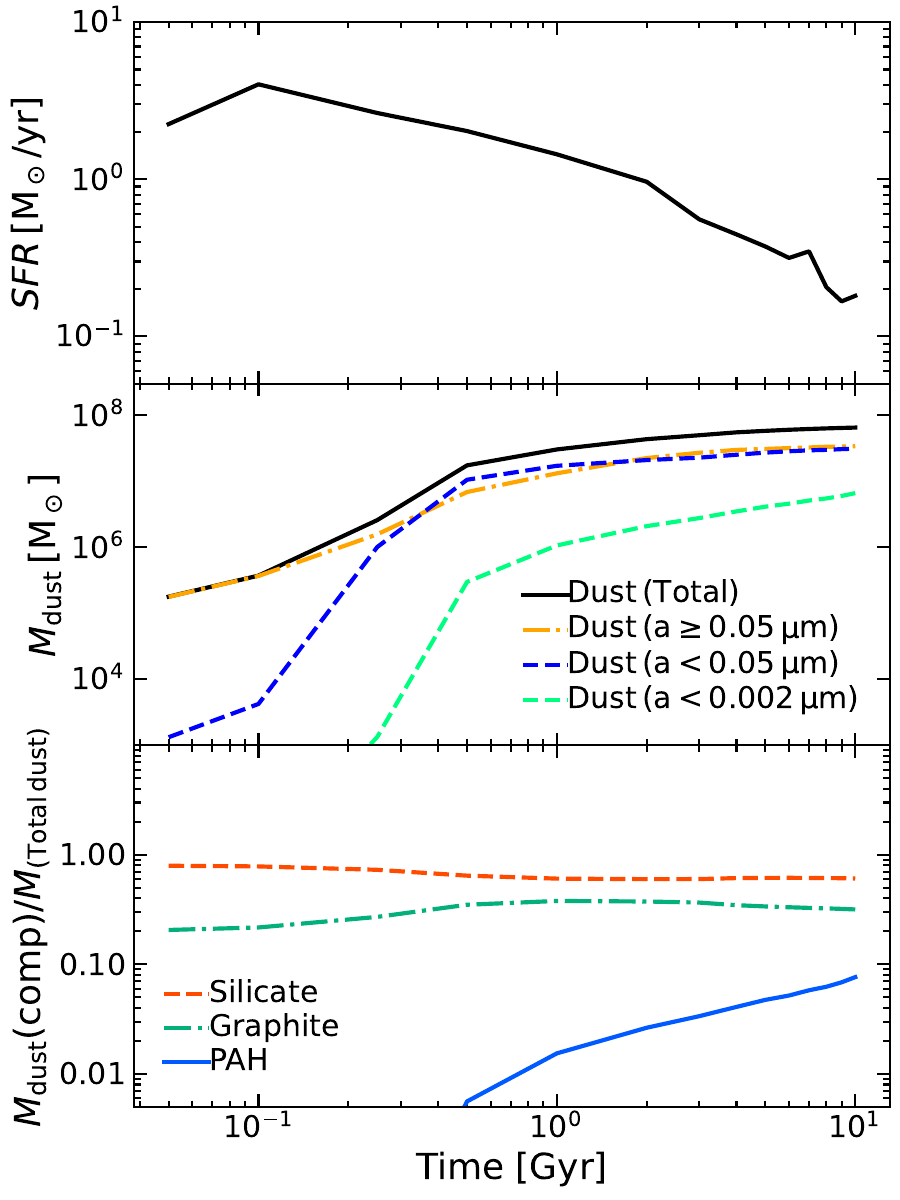}
    \caption{\textbf{Top panel:} star formation history of the MW-like galaxy simulation. \textbf{Middle panel:} evolution of the total dust mass over time, along with the dust masses of small grains ($a < 0.05\ \mathrm{\mu m}$) and very small grains ($a < 0.002\ \mathrm{\mu m}$). \textbf{Bottom panel:} time evolution of the dust masses for silicate, graphite, and PAHs. }%
    \label{fig: Global Evolution of the galactic properties}
\end{figure}
Figure~\ref{fig: Global Evolution of the galactic properties} shows the global evolution of the galactic properties of the MW-like galaxy simulation. The simulation shows a high initial star formation rate, which gradually declines over time, as shown in the top panel.
The middle panel presents the evolution of the total dust mass and the dust masses across different size ranges. The total dust mass increases over time mainly due to the accretion on dust grains.
Initially, dust is dominated by large dust grains ($a \geq 0.05\ \mathrm{\mu m}$) due to the dust production of SNe II, while the small grains  ($a < 0.05\ \mathrm{\mu m}$) are gradually produced by the interplay of shattering and accretion over time. 
The very small dust grains ($a < 0.002\ \mathrm{\mu m}$) including PAHs, form later compared to the small dust grains.
The bottom panel shows the mass evolution of silicate, graphite, and PAHs within the MW-like galaxy simulation.
Initially, silicate grains produced by SNe II dominate the dust mass in the galaxy. At later times, SNe Ia contribute additional gas-phase carbon, increasing the abundance of carbonaceous grains \citep{Saitoh2017}. After $t = 0.5$ Gyr, PAHs are produced through shattering, resulting in a dramatic increase in their mass fraction.
We note that our simulation does not take into account condensation efficiencies or accretion efficiencies that depend on the metal species.
These factors could potentially affect the mass evolution of the dust compositions \citep{Li2021, Choban2022, Dubois2024}.

\section{Evolution of the average albedo of the MW-like galaxy simulation}\label{app: Global evolution of albedo of the MW galaxy simulation}
\begin{figure}[]
    \centering
    \includegraphics[width=0.5\textwidth]{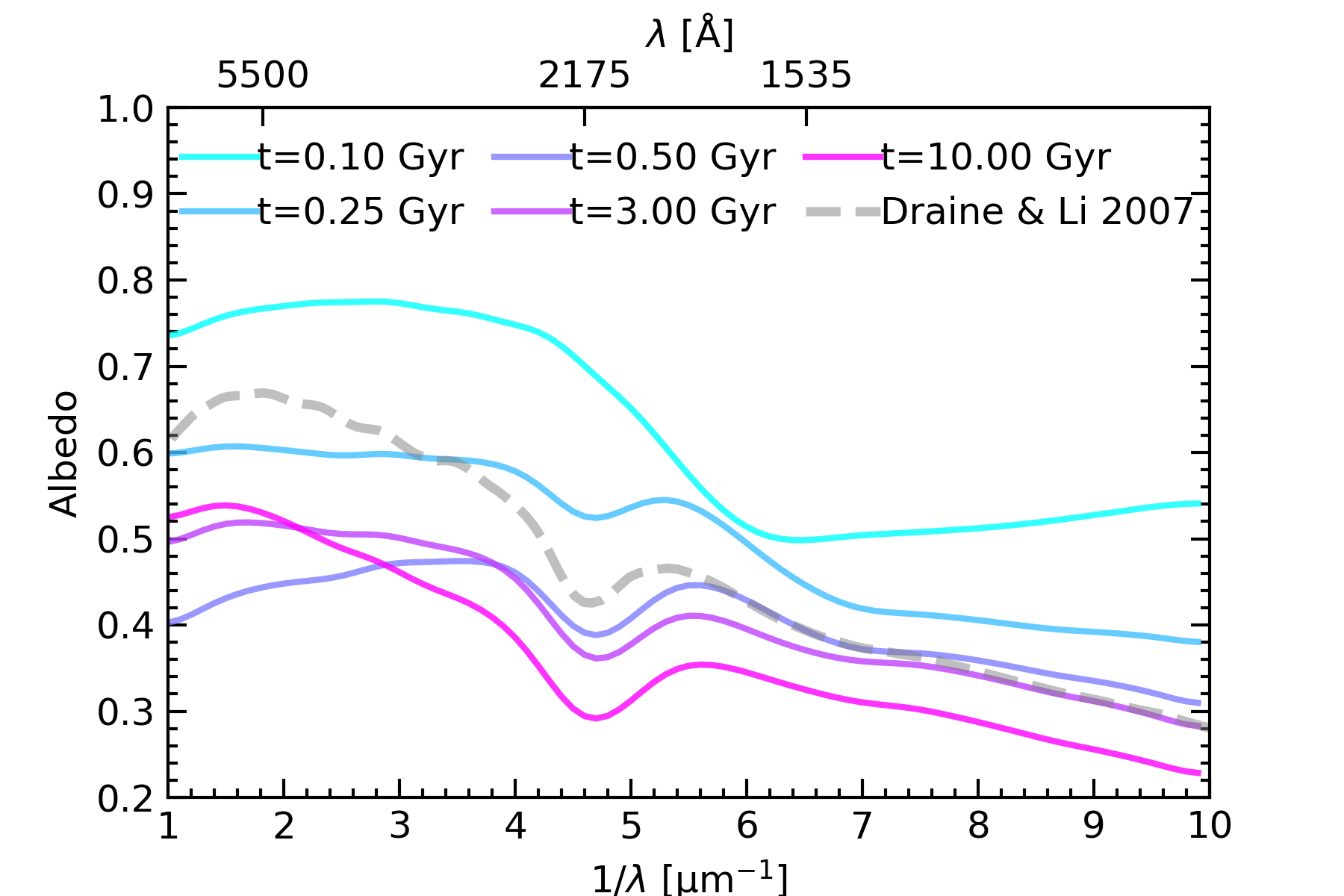}
    \caption{Evolution of the wavelength dependency of the average albedo of our MW-like galaxy simulation. The gray dashed line represent the wavelength dependency of the albedo of the \citet{Draine&Li2007} model. }%
    \label{fig: Global Evolution of the albedo}
\end{figure}

Figure~\ref{fig: Global Evolution of the albedo} shows the evolution of the wavelength dependency of the average albedo of the MW-like galaxy simulation. 
Overall, the average albedo at UV wavelengths is lower compared to optical wavelengths, since small grains with lower albedo are responsible for the opacity at UV wavelengths.
At $t = 0.10$ Gyr, large grains dominate in the galaxy, resulting in a relatively high albedo across all wavelengths.
Later, small grains form through shattering and accretion, and albedo gradually becomes lower up to $t=0.5$ Gyr. At the same time, the albedo at $2175 \ \AA$ becomes lower compared to other wavelengths.
After $t=0.5$ Gyr, the albedo at optical wavelengths becomes higher due to the formation of larger grains via coagulation, while the albedo at UV wavelengths continuously goes down due to the formation of very small grains including PAH through shattering. At $t=10$ Gyr, the albedo at all wavelengths is lower than that of \citet{Draine&Li2007} model.

\section{Evolution of global dust attenuation the MW-like galaxy simulation}\label{app: Evolution of dust attenuation}
\begin{figure*}[]
    \centering
    \includegraphics[width=0.95\textwidth]{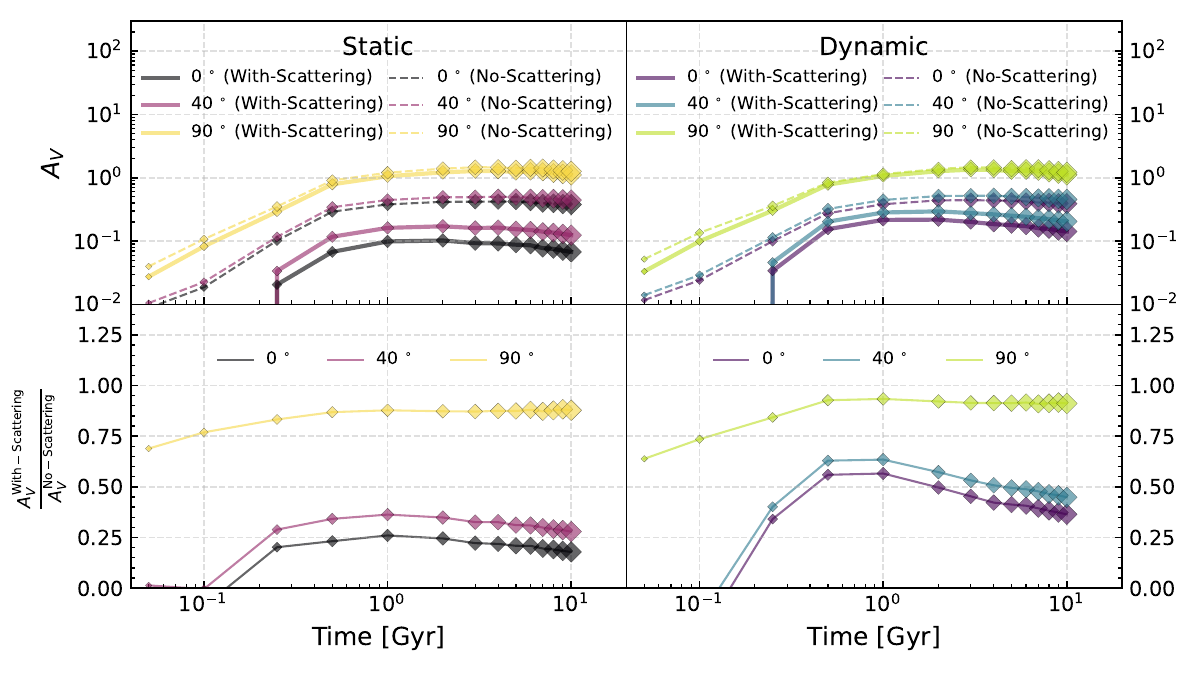}
    \caption{Time evolution of the global $A_V$ for the \textit{Static} and \textit{Dynamic} dust models in our Milky Way-like galaxy simulation is shown in the left and right panels, respectively. In each panel, solid lines indicate attenuation including scattering (\textit{With-Scattering}), while dashed lines represent attenuation neglecting scattering (\textit{No-Scattering}). The inclination angle is color-coded. The bottom panels show the ratio between the \textit{With-Scattering} and \textit{No-Scattering} attenuation. This ratio quantifies the impact of scattering on the total attenuation, with lower values indicating a stronger contribution from scattering.
    }%
    \label{fig: the time evolution of Av}
\end{figure*}
\begin{figure*}[]
    \centering
    \includegraphics[width=0.95\textwidth]{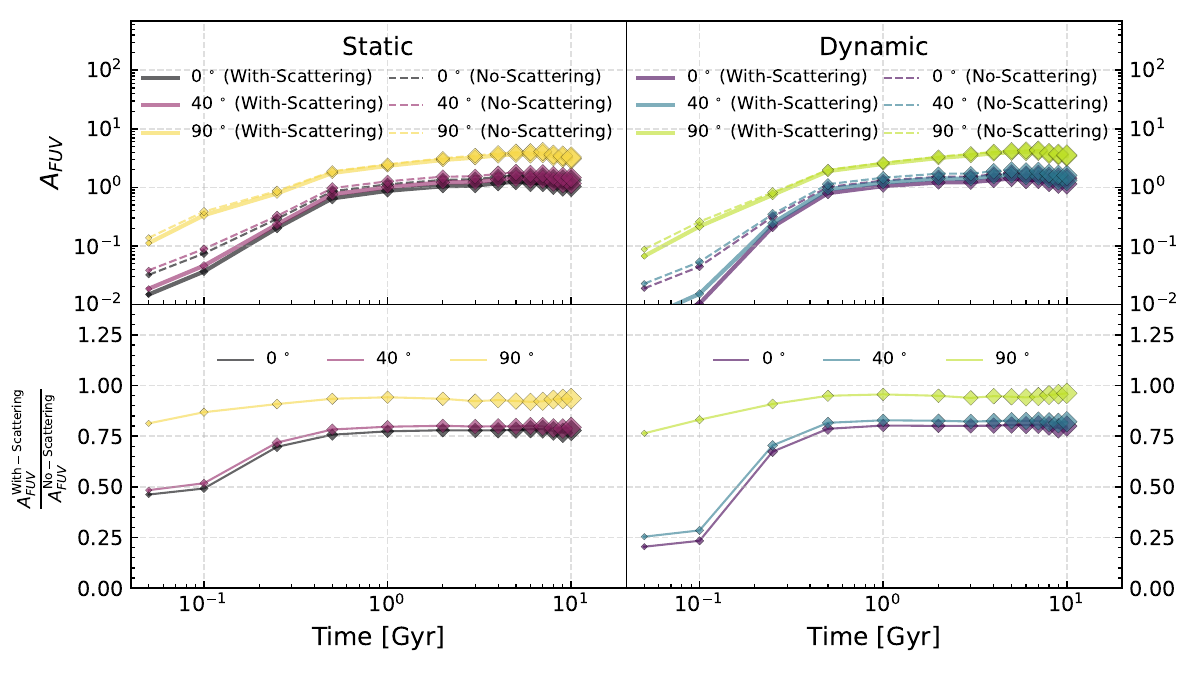}
    \caption{Same as Fig.~\ref{fig: the time evolution of Av}, but time evolution of the $A_{FUV}$.}%
    \label{fig: the time evolution of AFUV}
\end{figure*}

The upper left panel of Figure~\ref{fig: the time evolution of Av} shows the time evolution of the global $A_V$ for the static dust model, comparing results that include scattering (\textit{With-Scattering}) and those that neglect it (\textit{No-Scattering}).
At all inclination angles, $A_V$ increases over time as the total dust mass in the galaxy grows.
At an inclination angle of $90^\circ$, however, $A_V$ shows a slight decline after $t = 3$ Gyr for both the \textit{Static With-Scattering} and \textit{No-Scattering} models.
This decline is attributed to the decreasing fraction of young stars over time \citep{Narayanan2018}.
In contrast, at an inclination angle of $0^\circ$, $A_V$ for the \textit{Static With-Scattering} model begins to decrease after $t = 2$ Gyr, whereas $A_V$ for the \textit{Static No-Scattering} model continues to increase. This different behavior is driven by the effect of scattering.
The lower left panel of Figure~\ref{fig: the time evolution of Av} presents the time evolution of the ratio between $A_V$ in the \textit{Static With-Scattering} and \textit{Static No-Scattering} models, providing a quantitative measure of scattering effects on $A_V$.
At all inclination angles, this ratio increases over time, indicating that the impact of scattering is gradually reduced as the optical depth along the line of sight becomes larger. At an inclination angle of $90^\circ$, the ratio is close to unity, suggesting that scattering has a minimal effect.
In contrast, at lower inclination angles, the ratios are smaller, meaning that scattering reduces the observed $A_V$. Moreover, after $t = 1$ Gyr, the ratio declines further over time, implying that the contribution from scattering becomes increasingly significant because the dust disk geometry becomes more extended and scattering events become more likely.

The upper right panel of Figure~\ref{fig: the time evolution of Av} shows the time evolution of the global $A_V$ for the dynamic dust model.
The overall trends of $A_V$ for both the \textit{Dynamic With-Scattering} and \textit{No-Scattering} models are similar to those for the static models.
However, at lower inclination angles, $A_V$ for the \textit{Dynamic With-Scattering} model is higher than that of the \textit{Static With-Scattering} model. This is because large silicate grains ($a > 0.05 \ \mathrm{\mu m}$) are more abundant in the static model \citep{Draine&Li2007}, leading to enhanced scattering and consequently lower $A_V$.
The lower right panel of Figure~\ref{fig: the time evolution of Av} shows the time evolution of the ratio between $A_V$ in the \textit{Dynamic With-Scattering} and \textit{No-Scattering} models.
At all inclination angles, the ratios for the dynamic model increase more significantly than those for the static model up to $t = 0.5$ Gyr, driven by the formation of small grains with low albedo.
At lower inclination angles, the dynamic model exhibits higher ratios than the static model, indicating that scattering is less effective in reducing $A_V$.
While the ratio remains nearly constant at an inclination angle of $90\,^\circ$ after $t = 1$ Gyr, it declines at lower inclination angles. This trend is driven by the growth of large grains through coagulation in the ISM, which enhances scattering and further reduces $A_V$.

The upper left panel of Figure~\ref{fig: the time evolution of AFUV} shows the time evolution of the global $A_{FUV}$ for the \textit{Static With-Scattering} and \textit{No-Scattering} models.
At all inclination angles, $A_{FUV}$ for both the \textit{Static With-Scattering} and \textit{No-Scattering} models increases over time as the total dust mass in the galaxy grows up to $t = 7$ Gyr. At later times, $A_{FUV}$ declines because the intrinsic UV spectrum becomes dominated by old stars, reducing the contrast between the intrinsic and observed spectra.
The lower left panel of Figure~\ref{fig: the time evolution of AFUV} presents the time evolution of the ratio between $A_{FUV}$ in the \textit{Static With-Scattering} and \textit{Static No-Scattering} models.
Overall, the ratio at the FUV wavelength is larger than that at the V-band wavelength, indicating that scattering has a less significant effect at the FUV wavelength. This is primarily because the opacity at the FUV wavelength is generally higher than at the V-band wavelength, causing most FUV photons to be eventually absorbed.
The ratio at FUV wavelengths increases over time until $t=1.0$ Gyr, since the effect of scattering is reduced as the optical depth along the line of sight becomes larger. After $t=1$ Gyr, the ratio at the FUV wavelength remains constant.

The upper right panel of Figure~\ref{fig: the time evolution of AFUV} shows the time evolution of the global $A_{FUV}$ for the dynamic dust model.
The overall trends of $A_{FUV}$ for both the \textit{Dynamic With-Scattering} and \textit{No-Scattering} models are similar to those for the static models.
However, $A_{FUV}$ for the dynamic model starts with a lower value compared to the static models, since only large grains, which have low opacity at FUV wavelengths, are present at the beginning of the simulation. As small grains form through accretion and shattering over time, $A_{FUV}$ for the dynamic model becomes comparable to that for the static model.
The lower right panel of Figure~\ref{fig: the time evolution of AFUV} shows the time evolution of the ratio between $A_{FUV}$ in the \textit{Dynamic With-Scattering} and \textit{No-Scattering} models.
At all inclination angles, the ratios for the dynamic model increase more significantly than those for the static model up to $t = 0.5$ Gyr, driven by the formation of small grains with low albedo at the FUV wavelength.
After $t = 1.0$ Gyr, the ratio at the FUV wavelength remains constant.

In summary, $A_V$ is significantly influenced by scattering, particularly at lower inclination angles. The scattering processes are sensitive to changes in the star–dust geometry, such as inclination angle and the configuration of dust optical depths, as well as to variations in albedo driven by the evolving grain size distribution. In contrast, $A_{FUV}$ is less influenced by scattering due to its generally higher optical depths. Instead, it is more sensitive to the abundance of small grains and the star–dust geometry.

\section{Maps of attenuation at various wavelengths} \label{app: Maps of attenuation at various wavelengths}
\begin{figure*}[]
    \centering
    \includegraphics[width=0.95\textwidth]{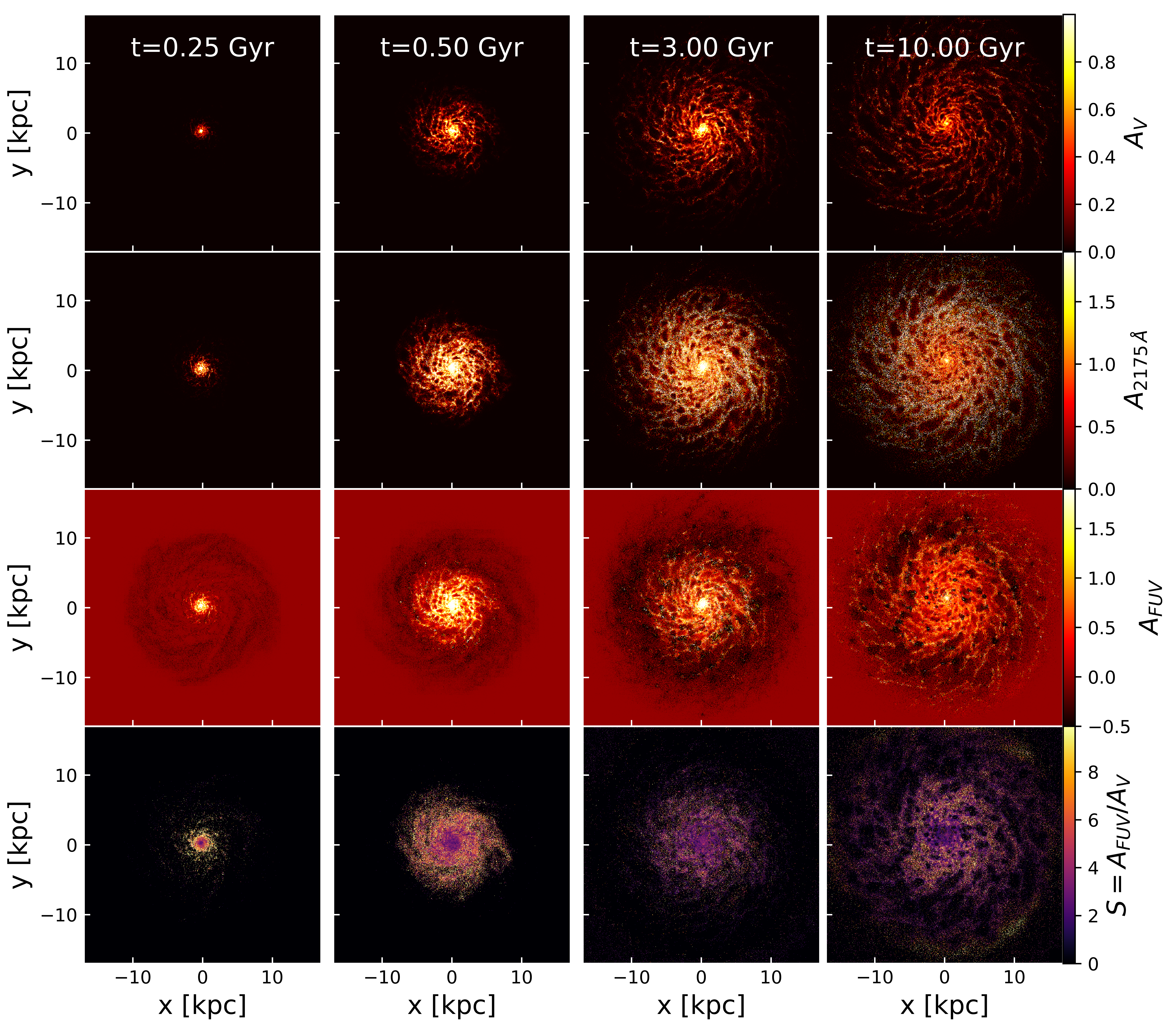}
    \caption{Time evolution of attenuation maps at $V$-band, $2175$ \AA{}, and $FUV$-band wavelengths as well as the map of slope (top to bottom panels). The time evolution is exhibited from the left to right panels.}%
    \label{fig: Band maps at i00}
\end{figure*}

The top panels of Fig.~\ref{fig: Band maps at i00} show the time evolution of the $A_V$ map viewed face-on. Initially, dust is produced by SNe II primarily in the central region of the galaxy. As time progresses, dust forms via accretion, extending outward along the spiral arms from the center to the outskirts. Large grains are responsible for $A_V$, and thus, the $A_V$ maps exhibit thin spiral arms tracing the dense gas in the ISM. 
The second panels from the top in Fig.~\ref{fig: Band maps at i00} show the time evolution of the attenuation maps at $\lambda=2175$ \AA{} ($A_{2175\, \AA}$), where the attenuation is primarily caused by small grains. Small grains form via shattering, especially in the diffuse regions surrounding the spiral arms. As a result, the $A_{2175\, \AA}$ maps exhibit thicker spiral arms.
The third panels from the top in Fig.~\ref{fig: Band maps at i00} present the time evolution of the $A_{FUV}$ map. Overall, its morphology is similar to that of the $A_{2175\, \AA}$ maps. At FUV wavelengths, scattering can be significant in regions with low dust surface density, allowing scattered photons to escape from the regions. As a result, $A_{FUV}$ becomes negative in such regions, particularly where supernova feedback sweeps out gas and dust. The bottom panels in Fig.~\ref{fig: Band maps at i00} show the time evolution of the slope maps. Overall, the slopes are flatter in the central regions of the galaxy compared to the outskirts. This is because dust in the center is optically thick to stellar emission at optical-to-UV wavelengths, suppressing scattering. In contrast, dust is optically thin at optical wavelengths in the outer regions, allowing scattered optical photons to escape and resulting in steeper slopes.

\end{appendix}

\end{document}